\documentclass[10pt, journal]{IEEEtran}
\usepackage[utf8x]{inputenc} 
\usepackage{amsmath}
\usepackage{amsfonts}
\usepackage{amssymb}
\usepackage{paralist}
\usepackage{graphicx}
\usepackage{booktabs}
\usepackage{bm}
\usepackage{algpseudocode}
\usepackage{multicol}
\usepackage{stfloats}
\usepackage{url}
\usepackage{enumitem}
\usepackage[table]{xcolor}
\usepackage{threeparttable}
\usepackage{multirow}
\usepackage[caption=false,font=footnotesize]{subfig}
\usepackage[level]{datetime}  
\newdateformat{ukdate}{\ordinaldate{\THEDAY} \monthname[\THEMONTH] \THEYEAR}
\usepackage[colorlinks]{hyperref}
\usepackage[nameinlink,noabbrev]{cleveref}
\usepackage[square, comma, sort&compress, numbers]{natbib}

\usepackage{epstopdf}
\usepackage{algorithm}
\usepackage{algorithmicx}
\usepackage{verbatim}                           
\usepackage{setspace}
\usepackage{soul}

\soulregister\cite7               
\soulregister\citep7              
\soulregister\citet7              
\soulregister\ref7                
\soulregister\pageref7            
\usepackage{color}
\graphicspath{{figures/}}
\IEEEaftertitletext{\vspace{-1.5\baselineskip}}
\usepackage{amsthm}

\makeatletter

\begin{document}

\title{Performance Analysis of uRLLC in scalable Cell-free Radio Access Network System}

\author{Ziyang Zhang$^{\ast\dagger}$, Dongming Wang$^{\ast\dagger}$, Yunxiang Guo$^{\ast\dagger}$, Yang Cao$^{\ast\dagger}$ and Xiaohu You$^{\ast\dagger}$\\
{$^\ast$National Mobile Communications Research Laboratory, Southeast University, Nanjing 210096, China}\\
{$^\dagger$Purple Mountain Laboratories, Nanjing 211111, China}\\
{Emails:\{ziyangzhang, wangdm, ieguoyunxiang, epstwxv, xhyu\}@seu.edu.cn}

}

\maketitle

\begin{abstract}
  As a critical component of beyond fifth-generation (B5G) and sixth-generation (6G) mobile communication systems, ultra-reliable low-latency communication (uRLLC) imposes stringent requirements on latency and reliability.
  In recent years, with the improvement of mobile communication network, centralized and distributed processing schemes for cell-free massive multiple-input multiple-output (CF-mMIMO) have attracted significant research attention.
  This paper investigates the performance of a novel scalable cell-free radio access network (CF-RAN) architecture featuring multiple edge distributed units (EDUs) under the finite block length regime.
  Closed expressions for the upper and lower bounds of its expected spectral efficiency (SE) performance are derived, where centralized and fully distributed deployment can be treated as two special cases, respectively.
  Furthermore, the spatial distribution of user equipments (UEs) and remote radio units (RRUs) is examined and the analysis reveals that the interleaving RRUs  
  deployment associated with the EDU can enhance SE performance under finite block length constraints with specific transmission error probability.
  The paper also compares Monte Carlo simulation results with multi-RRU clustering-based collaborative processing, validating the accuracy of the space-time exchange theory in the scalable CF-RAN scenario.
  By deploying scalable EDUs, a practical trade-off between latency and reliability can be achieved through spatial degree-of-freedom (DoF), offering a distributed and scalable realization of the space-time exchange theory.
\end{abstract}


\IEEEpeerreviewmaketitle

\section{INTRODUCTION}
\IEEEPARstart{W}{ith}
the continuous development of communication technology and the increasing demand for data transmission, the fifth-generation mobile communication system (5G) has been widely promoted. The main characteristics of 5G are higher data transmission speed, lower transmission latency, higher reliability, and a significantly higher number of connections compared to the fourth generation systems (4G).
Specifically, 5G regards low latency and high-reliability communication (uRLLC) as one of the three essential application scenarios \cite{dogra2020survey}.

URLLC must meet the requirements for transmission reliability and low latency. The work around uRLLC is mainly based on the research in \cite{polyanskiy2010channel} for the scenario of finite block length.
In finite block length, the traditional Shannon capacity based on the law of large numbers is no longer applicable \cite{xiaohu2020shannon}.
5G has adopted various technologies to achieve low latency and high reliability, among which is multi-transmit/receive points (multi-TRPs), one of the most important.
In the long-term evolution of the fourth-generation mobile communication system (4G-LTE), coordinated multi-point (CoMP) transmission is also used for multi-point transmission and reception. Multi-TRP and CoMP are critical technologies for improving spectrum efficiency (SE), peak rate, and reliability, which can collaboratively process signals between multiple base stations (BS), thereby improving network performance.
Compared to 5G, 6G requires higher reliability and lower transmission latency \cite{TowardTKU, wang2023road}, which requires more advanced technology.
In recent research work, \cite{park2020extreme} pointed out several critical limitations of uRLLC in the current 5G system and pointed out that scalability will be one of the key technical indicators for eXtreme ultra-reliable and low-latency communication (xURLLC) in 6G.
\cite{vision6gurllc} provided a comprehensive review of existing 5G uRLLC technology, shedding light on their associated risks and challenges in the context of future 6G communication systems.
Various types of interference challenges in uRLLC were discussed in \cite{B5GURLLC} and classified according to their deployment, design, technology, usage, and propagation characteristics, and extreme throughput performance under very low latency and up to $10^{-9}$ percentile success probability applications were evaluated. \cite{EfficientDe} provided a comprehensive review and comparison of different candidate decoding techniques for uRLLC regarding their error-rate performance and computational complexity for structured and random short block length.

Utilizing the abundant spatial resources of 6G networks has also become a research direction for URLLC implementation.
The concept of spatiotemporal exchangeability was novelly proposed in \cite{you20236g}, and a specific implementation scheme was proposed in \cite{you2022spatiotemporal}.
In the $15$\rm{dB} signal-to-noise ratio (SNR) scenario, the capacity collapse effect caused by finite block length can be compensated by increasing the number of streams in the system with the deployment in the spatial domain.
By appropriately selecting the code rate, block length, and the number of codewords in the time and spatial domains, the coding scheme proposed in \cite{you2022spatiotemporal} can achieve a good tradeoff between transmission latency and reliability.
In subsequent work, \cite{you2023closed} proved the exchangeability theory, deriving a closed expression for channel dispersion in massive multiple-input multiple-output (mMIMO).
Compact and explicit performance bounds of finite block length coded MIMO were formed to explore the relationship between block length, decoding error probability, rate, and DoF in different coding modes \cite{ye2024explicit}.
The above point-to-point work thoroughly and rigorously verified the idea of increasing the spatial degree-of-freedom (DoF) to compensate for the shortcomings of finite block length.

In addition, cell-free mMIMO (CF-mMIMO) is also one of the key technologies for implementing URLLC.
CF-mMIMO is a more revolutionary technology that can break the traditional cellular architecture and achieve cell-free communication with more flexible and efficient network coverage \cite{ngo2017cell, riera2018clustered}.
Researching how to further utilize cell-free architecture to implement uRLLC in 6G systems is worthwhile.
There have been many studies on cell-free systems to realize uRLLC.
\cite{you2021distributed} pointed out that cell-free architecture with a large number of distributed antennas has macro-diversity and spatial sparsity characteristics, which can further improve the performance of uRLLC.
In \cite{peng2022resource} and \cite{peng2023resource}, using successive convex approximation, the non-convex problem was transformed into a series of sub-problems for processing based on different precoding and combining schemes and , the quality of uRLLC services will benefit by deploying more access points (APs).
\cite{bucci2023performance} investigated the potential of CFmMIMO to multiplex uRLLC and other services by exploiting the sole spatial diversity through network slicing and adopting greedy pilot allocation to minimize pilot contamination.
A particular type of conjugate beamforming was proposed in \cite{nasir2021cell} that only required local CSI, and a new path-following algorithm was developed to optimize uRLLC rates and CF-mMIMO energy efficiency.
\cite{lancho2023cell} proposed a general framework to characterize the achievable grouping error probability in the CF-mMIMO based on saddlepoint approximation and scaling-based random coding union and the performance of CF-mMIMO supporting uRLLC is analyzed.
\cite{peng2023pilot} investigated the resource allocation problem of CF-mMIMO assisted uRLLC systems, proposed a new pilot allocation scheme that balances the ratio of pilot length to payload, and jointly optimized pilot and payload power, balancing estimated channel gain and pilot contamination.

In the deployment of CF-mMIMO supporting uRLLC, some challenges still need to be solved.
Firstly, for multi-user scenario processing, traditional centralized processing complexity will be very high so the cell-free deployment must face the problem of scalability.
Secondly, more efficient architecture deployment and advanced association algorithms are needed to support the complexity of distributed cell-free systems. 
The research on CF-mMIMO is still in the theoretical stage.
According to the scalability, \cite{bjornson2019making} introduced the classification method of CF-mMIMO with four different implementation levels.
\cite{cao2023oran} investigated the cell-free radio access network (CF-RAN) under the O-RAN architecture.
\cite{wang2023full} proposed a scalable architecture for CF-mMIMO systems through distributed transceivers and scalable collaborative transmission, which can further improve the network's performance. Based on this, potential vital technologies such as channel information acquisition, transceiver design, dynamic user and access point association, and new duplex were introduced, and the performance of distributed receiver design was evaluated.

However, we have noticed several shortcomings in the current research:
 (1) Existing work on the cell-free implementation of uRLLC mostly failed to consider the scalability and implementability of the system; 
 (2) Considering the actual deployment of cell-free APs, many studies analyzed the SE based on stochastic geometry architecture; The precoding and combining schemes have not fully utilized the collaborative characteristics of cell-free, and there is little research on collaborative transceivers based on interference suppression; 
 (3) The research on scalable cell-free was nearly all based on infinite block length, without considering the impact of finite block length on system performance, which cannot meet the uRLLC requirement.

 Therefore, in response to the above issues, relying on the currently validated spatiotemporal exchangeability theory \cite{you2023closed} in point-to-point transmission, this paper will implement uRLLC for scalable cell-free RAN systems. The main contributions of this paper are as follows:
\begin{itemize}
\item {The expected SE of a new scalable cell-free RAN with multiple edge distributed units (EDUs) is analyzed under finite block length.}
\item {A improved graph coloring algorithm for interleaving deployment is used to analyze the correlation performance of remote radio units (RRUs) under multiple EDUs that can improve the system SE under latency and reliability constraints.}
\item {By deploying scalable EDUs, a compromise between reliability and latency is exchanged with spatial DoF, further expanding and verifying the accuracy of the distributed space-time exchangeability theory.}
\end{itemize}

\textbf{Organization}: The rest of the paper is organized as follows.
In Section II, we detail the system model.
Section III analyzes the upper and lower bounds of expected SE with finite block length and then suppose the improved graph coloring algorithm to associate RRUs with EDUs.
Section IV specifies the simulation performance. Finally, Section V concludes this paper.

\textbf{Notation}: bold uppercase $\mathbf{A}$ (bold lowercase $\mathbf{a}$) denotes a matrix (a vector).
${{\mathbf{I}}_N}$ and ${{\mathbf{0}}_{M,N}}$ denote the $N \times N$ dimensional identity matrix and the $M \times N$ dimensional all-zero matrix, respectively.
${\left(  \cdot  \right)^{\rm{H}}}$, ${\left(  \cdot  \right)^{\rm{T}}}$ , ${\left(  \cdot  \right)^ * }$, ${\left(  \cdot  \right)^{-1} }$ and ${\left(  \cdot  \right)^ {\dagger} }$ stand for the conjugate transpose, transpose, conjugate, inverse and pseudo-inverse, respectively.
${\mathrm{diag}}\left\{ {\mathbf{a}} \right\}$, ${\mathrm{diag}}\left\{ {\mathbf{A}} \right\}$ and ${\mathrm{blkdiag}}\left\{ {\mathbf{A}_1}, \cdots, {\mathbf{A}_N} \right\}$ represent a diagonal matrix with $\mathbf{a}$ along its main diagonal, a vector constructed by the main diagonal of the matrix $\mathbf{A}$, a block diagonal matrix , respectively.
$\otimes$ denotes the Kronecker product of two matrices.
$\ell_0$ , $\ell_1$ and $\ell_2$ norm of vectors are denoted by $\left \| \cdot  \right \|_0$, $\left \| \cdot  \right \|_1$ and $\left \| \cdot  \right \|_2$, respectively. ${\mathcal{CN}}\left( \boldsymbol{\mu } , \mathbf{R} \right)$ denotes the complex Gaussian distribution with mean $\boldsymbol{\mu }$ and covariance matrix $\mathbf{R}$. ${\mathbb{E}}\left\{  \cdot  \right\}$ is the expectation operator.
Finally, $\backslash$ denotes the set subtraction operation.

\section{SYSTEM MODEL}

In this section, we introduce the implementation of the CF-mMIMO system.
Under the new architecture, we analyze the uplink SE of CF-mMIMO, and the combining strategy adopts a unified representation. We consider a CF-mMIMO with $LN$ antenna RRUs and $K$ single-antenna user equipments (UEs).
Assuming $LN$ is large, and {$L\gg K$}.
At the $l$-th RRU, the received signal ${{\bf{y}}_{{\rm{UL}},l}}$ can be expressed as \cite{bjornson2019making}

\begin{equation}
  \label{y_ul_01}
  {{\bf{y}}_{{\rm{UL}},l}} = \sum\limits_{k = 1}^K {{{\bf{h}}_{l,k}}\sqrt {{p_k}} {s_k}}  + {{\bf{z}}_l},
\end{equation}
where ${s_k}$ denotes the transmitted symbol of the $k$-th UE, ${p_k}$ denotes the uplink transmission power of the $k$-th UE, ${{\bf{h}}_{l,k}}$ represents the $N\times 1$ CSI from the $k$-th UE to the $l$-th RRU and ${{\bf{z}}_l} \sim {{\cal CN}}\left( {0,{\sigma_{\rm UL}^2}{{\bf{{I}}}_N}} \right)$ represents the additive white gaussian noise (AWGN).

\begin{figure}[!htb]
  \centering
  \includegraphics[width=3.5in]{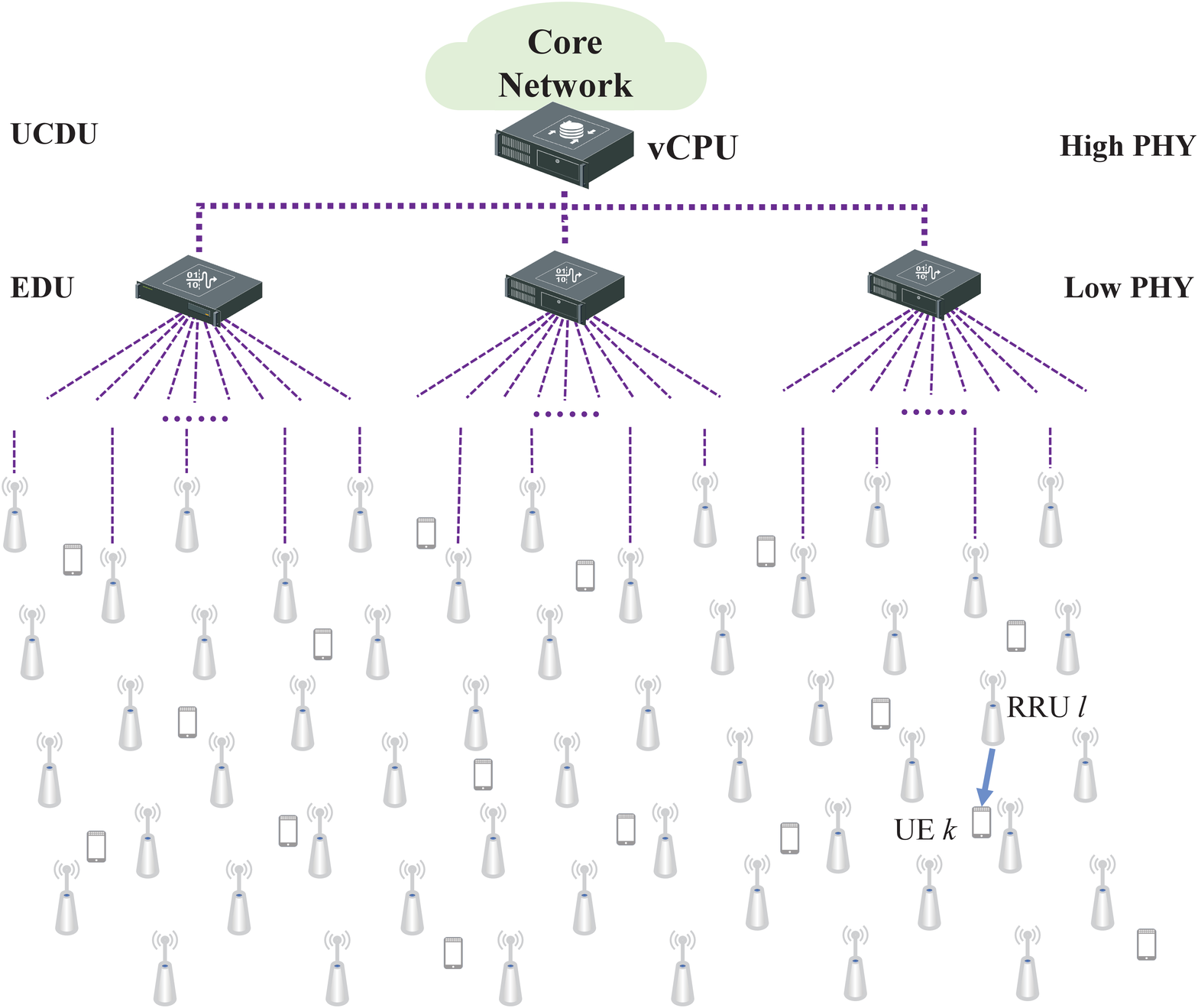}
  \caption{The system has $M$ EDUs, $L$ RRUs, each RRU equiped with $N$ antennas, and $K$ single-antenna UEs.}
  \label{fig1_system_model}
\end{figure}

We assume there are $M$ EDUs in the system connected to the user-centric distributed unit (UCDU), as shown in Fig. \ref{fig1_system_model}.
A splitting option between low-PHY and high-PHY is proposed in \cite{cao2023oran}, let the channel vector from the $k$-th UE to all RRUs be,

\begin{equation}
  \label{channel}
  {{\mathbf{h}}_{k}}={{\left[ \mathbf{h}_{k,1}^{\rm T},\mathbf{h}_{k,2}^{\rm T}\cdots \mathbf{h}_{k,L}^{\rm T} \right]}^{\rm T}}\in {{\mathbb{C}}^{LN}},
\end{equation}
among them, ${{\mathbf{h}}_{k}}={{{\mathbf{B }}^{1/2}_{k}}}{{\mathbf{g}}_{k}}$, the large-scale fading matrix ${{\mathbf{B }}_{k}}={\rm diag}\left( {{\beta  }_{k,1}},{{\beta  }_{k,2}},\cdots ,{{\beta  }_{k,L}} \right)\otimes {{\mathbf{I}}_{N}}$.
The correlated Rayleigh fading channel vector ${{\mathbf{g}}_{k}}$ is distributed as ${{\mathbf{g}}_{k}}\sim{{\mathcal{CN}}}\left( \mathbf{0},{{\mathbf{R}}_{k}} \right)$, where ${{\mathbf{R}}_{k}}=\operatorname{diag}\left( {{\mathbf{R}}_{k,1}},\ldots ,{{\mathbf{R}}_{k,L}} \right)\in {{\mathbb{C}}^{LN\times LN}}$ represents the block diagonal spatial correlation matrix of UE $k$.

The uplink signal-to-interference plus noise ratio (SINR) for the $k$-th UE is shown on the top of the next page,
\begin{figure*}[ht!]
  \label{SINR_ULDL}
  \begin{align}
    \gamma _k^{({\rm{UL}},{\rm{d}})} = \frac{{{p_k}{{\left| {\sum\limits_{m = 1}^M \mathbb{E} {\left( {{\bf{v}}_{k,m}^{\rm{H}}{{\bf{D}}_{k,m}}{{\bf{h}}_{k,m}}} \right)} } \right|}^2}}}{{\sum\limits_{i = 1,i \ne k}^K {{p_i}} \mathbb{E} {{\left| {\sum\limits_{m = 1}^M {{\bf{v}}_{k,m}^{\rm{H}}{{\bf{D}}_{k,m}}{{\bf{h}}_{i,m}}} } \right|}^2} + \sigma _{{\rm{UL}}}^2 \mathbb{E} \left( {\sum\limits_{m = 1}^M {{{\left\| {{{\bf{D}}_{k,m}}{{\bf{v}}_{k,m}}} \right\|}^2}}}\right)}},
  \end{align}
\hrulefill
\end{figure*}
where ${\bf{v}}_{k,m}$ is combining vector of UE $k$,
${{\bf{D}}_{k,m}}$ is the association matrix between the UE and the EDU, ${{\mathbf{D}}_{k}}={\rm diag}\left( {{\delta }_{k,1}},{{\delta }_{k,2}},\cdots ,{{\delta }_{k,L}} \right)\otimes {{\mathbf{I}}_{N}}$, where ${{\delta }_{k,l}}=1$, if RRU $l$ is associated with UE $k$, otherwise ${{\delta }_{k,l}}=0$.

For analysis, we assume perfect CSI with full association between UEs and RRUs. Considering a single-antenna RRU scenario, i.e., $N=1$, ${{\bf{D}}_{k,m}}={{\bf{I}}_{k,m}}$. Taking uplink reception as an example, the uplink reception SINR can be expressed in a general way as
\begin{equation}
  \label{uplink_ins_SINR}
  \gamma _k^{({\rm{UL}},{\rm{d}})} = \frac{{{p_k}{{\left| {\sum\limits_{m = 1}^M {{\bf{v}}_{k,m}^{\rm{H}}{{\bf{h}}_{k,m}}} } \right|}^2}}}{{\sum\limits_{i = 1,i \ne k}^K {{p_i}} {{\left| {\sum\limits_{m = 1}^M {{\bf{v}}_{k,m}^{\rm{H}}{{\bf{h}}_{i,m}}} } \right|}^2} + \sigma _{{\rm{UL}}}^2\left( {\sum\limits_{m = 1}^M {\left\|{{\bf{v}}_{k,m}}\right\|{^2}} } \right)}}.
\end{equation}

Clearly, centralized and fully distributed configurations can be introduced as exceptional cases for $M=1$ and $M=L$, respectively \cite{bjornson2020scalable}. In the next section, we take uplink analysis as an example and analyze the performance relationship between uplink SE and the distribution of RRU location.

Considering the impact of the finite block length, under AWGN channel conditions, given the required transmission error probability ${{\varepsilon}}$ and the transmission block length $n$, it is known that using non-Gaussian codebooks can achieve the maximum channel rate\cite{polyanskiy2010channel} and under finite block length, the SE of the $k$-th UE can be closely approximated as

\begin{equation}
  \label{rate}
  {{R}_k}\approx {{\log }_{2}}\left( 1+{{\gamma }_k} \right)-\sqrt{\frac{{{V}_k}}{n}}{{Q}^{-1}}\left( {{\varepsilon }_k} \right),
\end{equation}
where the channel dispersion term is

\begin{equation}
  {{V}_k}\text{=}\left( 1-{{\left( 1+{{\gamma }_k} \right)}^{-2}} \right){{\log }_{2}}^{2}e,
\end{equation}
in which ${{Q}^{-1}}$is the inverse of $Q(x)=\frac{1}{\sqrt{2\pi }}\int_{x}^{\infty }{{{e}^{-\frac{{{t}^{2}}}{2}}}}\text{d}t$.

Considering the independence of the equivalent cell-free channel, similar to the treatment in \cite{yang2014quasi}, the system SE can be expressed as

\begin{equation}
  \label{sum_rate}
  \begin{aligned}
    R &= C - \sqrt {\frac{V}{n}} {Q^{ - 1}}\left( \varepsilon  \right) \\
    &= \sum\limits_{i=1}^{K}{{{\log }_{2}}\left( 1+{{\gamma }_{i}} \right)}-\sqrt{\frac{K-\sum\limits_{i=1}^{K}{\frac{1}{{{\left( 1+{{\gamma }_{i}} \right)}^{2}}}}}{n}}{{Q}^{-1}}\left( {{\varepsilon }} \right),   
  \end{aligned}
\end{equation}
where $C$ is the traditional channel capacity and $V$ is the total channel dispersion parameter under multiple UEs.

\textit{Remark 1: Considering the multi-user interference, the SE analysis of UEs under finite block length becomes very complex. By adopting cooperative combining schemes with interference suppression, interference can be effectively mitigated, especially when the channel estimation is perfect, thus approximating the channel condition as an AWGN channel.
This approach is adopted in many current research works \cite{schiessl2019delay,ren2020joint}. The following section will analyze the system's SE with the ZF combining scheme for theoretical rigor.
}

\section{SE Analysis with Finite block length}

\subsection{Upper and Lower Bounds of SE}

For the ZF combining of perfect CSI \cite{zhang2021local}, the original centralized combining matrix is ${\bf{V}} = {\bf{H}}{\left( {{{\bf{H}}^{\rm{H}}}{\bf{H}}} \right)^{ - 1}} \in {\mathbb{C} ^{K \times LN}}$ assumed that the RRU is single-antenna. Therefore, ${\bf{H}} \buildrel \Delta \over = [{{\bf{h}}_1}, \ldots ,{{\bf{h}}_k}, \ldots ,{{\bf{h}}_K}] \in {\mathbb{C} ^{L \times K}}$. Since ZF completely nullifies interference, the interference received by each EDU is $0$. As the number of EDUs increases, the matrix dimensions continually decrease and we have:

\begin{equation}
  \label{uplink_ZF_SINR}
  \begin{aligned}
  \gamma _k^{({\rm{UL}},{\rm{d}})} &= \frac{{{p_k}{{\left| {\sum\limits_{m = 1}^M {{\bf{v}}_{k,m}^{\rm{H}}{{\bf{h}}_{k,m}}} } \right|}^2}}}{{\sum\limits_{i = 1,i \ne k}^K {{p_i}} {{\left| {\sum\limits_{m = 1}^M {{\bf{v}}_{k,m}^{\rm{H}}{{\bf{h}}_{i,m}}} } \right|}^2} + \sigma _{{\rm{UL}}}^2\left( {\sum\limits_{m = 1}^M {{{\left\| {{{\bf{v}}_{k,m}}} \right\|}^2}} } \right)}}\; \\
  &= \frac{{{p_k}{M^2}}}{{\sum\limits_{m = 1}^M {{{\left\| {{{\bf{v}}_{k,m}}} \right\|}^2}} }},
  \end{aligned}
\end{equation}
where
\begin{equation}
  \label{uplink_ZF}
  {{\bf{V}}_m} = {{\bf{H}}_m}{\left( {{\bf{H}}_m^{\rm{H}}{{\bf{H}}_m}} \right)^{ - 1}},
\end{equation}
${{\bf{v}}_{k,m}}$ is the $k$-th column of ${{\bf{V}}_m}$. Here, the condition for precoding requires that the number of antennas in the EDU ${L_m}$ is much greater than the number of UE $K$, and the UE set is ${\cal K}$. {Therefore, the total average SE of the system is shown on the top of the next page,
\begin{figure*}[ht!]
  \begin{equation}
    \label{uplink_sum_Rate}
    R=\sum\limits_{i=1}^{K}{{{\log }_{2}}\left( 1+\frac{{{p}_{k}}{{M}^{2}}}{\sum\limits_{m=1}^{M}{{{\left\| {{\mathbf{v}}_{k,m}} \right\|}^{2}}}} \right)}-\frac{{{Q}^{-1}}\left( \varepsilon  \right)}{\sqrt{n}}\sqrt{K-\sum\limits_{i=1}^{K}{\frac{1}{{{\left( 1+{{{p}_{k}}{{M}^{2}}}/{\sum\limits_{m=1}^{M}{{{\left\| {{\mathbf{v}}_{k,m}} \right\|}^{2}}}}\; \right)}^{2}}}}},
  \end{equation}
  \hrulefill
\end{figure*}

From (\ref{uplink_ZF}), we have that

\begin{equation}
  \label{uplink_Avg_ZF_SINR1}
  {{\left\| {{\mathbf{v}}_{k,m}} \right\|}^{2}} = \left[ {{{\left( {{\bf{H}}_m^{\rm{H}}{{\bf{H}}_m}} \right)}^{ - 1}}} \right]_{kk}.
\end{equation}

Among them, different channel matrices ${{\bf{H}}_1},{{\bf{H}}_2}, \ldots ,{{\bf{H}}_M}$ are independent, where ${{\bf{H}}_m} \in {\mathbb{C} ^{{L_m} \times K}}$.
The number of RRUs in an EDU satisfies the following three expressions,

\begin{subequations}
\label{group}
\begin{align}
  &\;\;\;{L_1} + {L_2} + \cdots + {L_M} = L,\\
  &\;\;\;{{\cal L}_1} \cup {{\cal L}_2} \cup \cdots \cup {{\cal L}_M} = {\cal L},\\
  &\;\;\;\left( {{\cal L}_1}\cap {{\cal L}_2} \right)\cup \cdots \cup \left( {{\cal L}_1}\cap {{\cal L}_M} \right) \cup \cdots \cup \left( {{\cal L}_{M-1}}\cap {{\cal L}_M} \right)=\varnothing,
\end{align}
\end{subequations}
where ${\cal L} = \{ l|\forall l = 1, \ldots ,L\}$. Clearly, the system's ergodic achievable sum SE can be expressed in the form of its expectation, i.e., $\overline R = {\mathbb{E}_{{{\bf{H}}_1},{{\bf{H}}_2}, \ldots ,{{\bf{H}}_M}}}\left\{ {\sum\limits_{i = 1}^K {{R_i}} } \right\}$.
We denote the left-hand side (LHS) of the ergodic achievable SE in (\ref{uplink_sum_Rate}) as $X = {\mathbb{E}_{{{\bf{H}}_1},{{\bf{H}}_2}, \ldots ,{{\bf{H}}_M}}}\left\{ {\sum\limits_{i = 1}^K {{{\log }_2}\left( {1 + \frac{{{p_k}{M^2}}}{{\sum\limits_{m = 1}^M {{{\left\| {{{\bf{v}}_{k,m}}} \right\|}^2}} }}} \right)} } \right\}$, and the right-hand side (RHS) of (\ref{uplink_sum_Rate}) as $Y = \frac{{{Q}^{-1}}\left( {{\varepsilon}} \right)}{\sqrt{n}}{{\mathbb{E}}_{{{\mathbf{H}}_{1}},{{\mathbf{H}}_{2}},\ldots ,{{\mathbf{H}}_{M}}}}\left\{ \sqrt{{{V}}} \right\}$.

Next, we will analyze them separately. For the LHS in (\ref{uplink_sum_Rate}), using Jensen's inequality, the upper bound can be expressed as

\begin{equation}
  \label{uplink_Avg_ZF_SINR_upper}
  X \le \sum\limits_{i = 1}^K {{{\log }_2}\left( {1 + {\mathbb{E}_{{{\bf{H}}_1},{{\bf{H}}_2}, \ldots ,{{\bf{H}}_M}}}\left\{ {\frac{{p_i}{M^2}}{{\sum\limits_{m = 1}^M {{\left[ {{{\left( {{\bf{H}}_m^{\rm{H}}{{\bf{H}}_m}} \right)}^{ - 1}}} \right]}_{kk}}}}}\right\}} \right)}.
\end{equation}

Similarly, the lower bound of the LHS can be expressed as,

\begin{equation}
  \label{uplink_Avg_ZF_SINR_lower}
  X \ge \sum\limits_{i = 1}^K {{{\log}_2}\left( {1 + \frac{{{p_i}{M^2}}}{{{\mathbb{E} _{{{\bf{H}}_1},{{\bf{H}}_2}, \ldots ,{{\bf{H}}_M}}}\left\{ {{\sum\limits_{m = 1}^M {{{\left[ {{{\left( {{\bf{H}}_m^{\rm{H}}{{\bf{H}}_m}} \right)}^{ - 1}}} \right]}_{kk}}} }}\right\}}}} \right)}.
\end{equation}

\textit{Remark 2: In comparison with the results obtained from centralized processing presented in \cite{liu2019spectral}, expressions (\ref{uplink_Avg_ZF_SINR_upper}) and (\ref{uplink_Avg_ZF_SINR_lower}) in our study compute the average SINR in the ergodic achievable rate by taking the expectations of the channel matrices for different EDUs independently. This approach utilizes the channel's independence between the EDUs and the UEs.}
}

Therefore, for the RHS of (\ref{uplink_sum_Rate}), using Jensen's inequality, we can have the same treatment, and the upper bound can be expressed as,

\begin{equation}
  \label{uplink_V_ZF_SINR_upper}
  Y \le \frac{{{Q}^{-1}}\left( {{\varepsilon}} \right)}{\sqrt{n}}\sqrt {K - \sum\limits_{i = 1}^K {1/{{\left( {1 + \frac{{{p_k}{M^2}}}{S}} \right)}^2}}},
\end{equation}
where $S = {{{\mathbb{E}_{{{\bf{H}}_1},{{\bf{H}}_2}, \ldots ,{{\bf{H}}_M}}}\left\{ {\sum\limits_{m = 1}^M {{{\left[ {{{\left( {{\bf{H}}_m^{\rm{H}}{{\bf{H}}_m}} \right)}^{ - 1}}} \right]}_{kk}}} } \right\}}}$.

Similarly, the lower bound of can be expressed as,
\begin{equation}
  \label{uplink_V_ZF_SINR_lower}
  Y \ge \frac{{{Q}^{-1}}\left( {{\varepsilon}} \right)}{\sqrt{n}}\sqrt {K - \sum\limits_{i = 1}^K {1/{{\left( {1 + {\mathbb{E}_{{{\bf{H}}_1},{{\bf{H}}_2}, \ldots ,{{\bf{H}}_M}}}\left\{T \right\}} \right)}^2}}}.
\end{equation}
where $T = {\frac{{{p_k}{M^2}}}{{\sum\limits_{m = 1}^M {{{\left[ {{{\left( {{\bf{H}}_m^{\rm{H}}{{\bf{H}}_m}} \right)}^{ - 1}}} \right]}_{kk}}}}}}$.

The fraction within the expectation can be analyzed using the matrix inversion formula,

\begin{equation}
  \label{Matrix_inverse_formular}
  {\left( {{\bf{H}}_m^{\rm{H}}{{\bf{H}}_m}} \right)^{ - 1}} = \frac{{{{\left( {{\bf{H}}_m^{\rm{H}}{{\bf{H}}_m}} \right)}^*}}}{{\det \left( {{\bf{H}}_m^{\rm{H}}{{\bf{H}}_m}} \right)}}.
\end{equation}

Let matrix ${\bf{G}}_{m}={{\bf{H}}_m^{\rm{H}}{{\bf{H}}_m}}$, ${\bf{G}}_{m,k}$ be the algebraic cofactor of the $(k,k)$-th element of matrix ${\bf{G}}_{m}$, so the $(k,k)$-th element of the inverse matrix ${{\bf{H}}_m^{\rm{H}}{{\bf{H}}_m}}$ is expressed as,

\begin{equation}
  \label{Matrix_inverse}
  {{{\left[ {{{\left( {{\bf{H}}_m^{\rm{H}}{{\bf{H}}_m}} \right)}^{ - 1}}} \right]}_{kk}}} = \frac{\det\left( {{{\bf{G}}_{m,k}}}\right)}{{\det \left( {{{\bf{G}}_m}} \right)}}.
\end{equation}

Therefore,
\begin{equation}
  \label{inverse_formular}
    \frac{1}{{\sum\limits_{m = 1}^M {{{\left\| {{{\bf{v}}_{k,m}}} \right\|}^2}} }} = \frac{1}{{\sum\limits_{m = 1}^M {{{\left[ {{{\left( {{\bf{H}}_m^{\rm{H}}{{\bf{H}}_m}} \right)}^{ - 1}}} \right]}_{kk}}} }}
    = \frac{1}{{\sum\limits_{m = 1}^M {{{\left[ \frac{\det\left( {{{\bf{G}}_{m,k}}}\right)}{{\det \left( {{{\bf{G}}_m}} \right)}} \right]}}}}}.
\end{equation}

Using Jensen's inequality, we can obtain:

\begin{equation}
  \label{inverse_formular_jensen}
  \begin{aligned}
    \frac{1}{{\frac{1}{M}\left[ {\sum\limits_{i = 1}^M {\frac{{\det \left( {{{\bf{G}}_{i,k}}} \right)}}{{\det \left( {{{\bf{G}}_i}} \right)}}} } \right]}} \le \frac{1}{M}\sum\limits_{i = 1}^M {\frac{1}{{\frac{{\det \left( {{{\bf{G}}_{i,k}}} \right)}}{{\det \left( {{{\bf{G}}_i}} \right)}}}}}  = \frac{1}{M}\sum\limits_{i = 1}^M {\frac{{\det \left( {{{\bf{G}}_i}} \right)}}{{\det \left( {{{\bf{G}}_{i,k}}} \right)}}}.
  \end{aligned}
\end{equation}

Therefore, according to (\ref{inverse_formular_jensen}), the upper bound of the LHS $X$ and the RHS $Y$ in (\ref{uplink_Avg_ZF_SINR_upper}) can be written on the top of next page.

\begin{figure*}[ht!]
\begin{equation}
  \label{uplink_X_upper}
  \begin{aligned}
  X \le {X^{{\rm{ub}}}}= \sum\limits_{i = 1}^K {{{\log }_2}\left( {1 + {p_i}M{\mathbb{E}_{{{\bf{H}}_1},{{\bf{H}}_2}, \ldots ,{{\bf{H}}_M}}}\left\{ {\frac{1}{M}\sum\limits_{m = 1}^M {\frac{{\det \left( {{{\bf{G}}_m}} \right)}}{{\det \left( {{{\bf{G}}_{m,i}}} \right)}}} } \right\}} \right)}.
  \end{aligned}
\end{equation}
\hrulefill
\end{figure*}

\begin{figure*}[ht!]
  \begin{equation}
    \label{uplink_Y_SINR_upper}
      Y \le {Y^{{\rm{ub}}}}= \frac{{{Q}^{-1}}\left( {{\varepsilon}} \right)}{\sqrt{n}}\sqrt {K - \sum\limits_{i = 1}^K {1/{{\left( {1 + {p_i}M{\mathbb{E}_{{{\bf{H}}_1},{{\bf{H}}_2}, \ldots ,{{\bf{H}}_M}}}\left\{ {\frac{1}{M}\sum\limits_{m = 1}^M {\frac{{\det \left( {{{\bf{G}}_m}} \right)}}{{\det \left( {{{\bf{G}}_{m,i}}} \right)}}} } \right\}} \right)}^2}\;}}.
  \end{equation}
\hrulefill
\end{figure*}

Using the Schur complement lemma \cite{wang2015asymptotic}, we have,
\begin{equation}
  \label{Schur_app}
  \begin{aligned}
    {\frac{{\det \left( {{{\bf{G}}_i}} \right)}}{{\det \left( {{{\bf{G}}_{i,k}}} \right)}}}
    & \mathop \approx \limits^{{L_m} \gg K} {\left\| {{{\bf{h}}_{k,m}}} \right\|^2} - \sum\limits_{j \ne k,j \in {\cal K}}  \cdot \frac{{{{\bf{h}}_{k,m}}} {\bf{h}}_{j,m}^{\rm{H}}{{\bf{h}}_{j,m}}{\bf{h}}_{k,m}^{\rm{H}}}{{{{\left\| {{{\bf{h}}_{j,m}}} \right\|}^2}}} \\
    &= {\left\| {{{\bf{h}}_{k,m}}} \right\|^2} - \sum\limits_{j \ne k,j \in {\cal K}} {{{\left| {{h_{k,l_{j,m}^*}}} \right|}^2}} = \sum\limits_{l \in {{\widetilde {\cal L}}_{m,k}}} {{{\left| {{h_{k,l}}} \right|}^2}},
  \end{aligned}
\end{equation}
where ${\widetilde {\cal L}_{m,k}} = {{\cal L}_m} - {\left\{ {l_{j,m}^*} \right\}_{j \in {\cal K},j \ne k}}$, $l_{j,m}^*$ represents the RRU closest to the $j$-th UE within the $m$-th EDU, and each EDU needs to perform this operation to chooses the closest RRU. According to the notation in the paper, define ${{\cal A}_{m,k}} \buildrel \Delta \over = {\rm{Unique}}\left( {\left\{ {l_{m,n}^ \star  = \arg \mathop {\max }\limits_{l \in {{\cal L}_m}} {\beta _{l,n}}|\forall n \ne k} \right\}} \right)$, and ${\widetilde {\cal L}_{m,k}}$ can be expressed as ${\widetilde {\cal L}_{m,k}} = {{\cal L}_m}/{{\cal A}_{m,k}}$.

{\textbf{Lemma 1}}: If the random variable $Y_i$ follows independent Gamma distributions, i.e., ${Y_i}\sim \Gamma \left( {{k_i},{\theta _i}} \right)$, then ${f_{{Y_i}}}\left( y \right) = \frac{1}{{{\theta _i}^{{k_i}}\Gamma \left( {{k_i}} \right)}}{y^{{k_i} - 1}}{e^{ - \frac{y}{{{\theta _i}}}}}$, where \cite{zhang2017downlink} provides the Gamma approximation expression. The first and second moments of the sum of multiple Gamma-distributed random variables $Y_i$ satisfy:
\begin{subequations}
  \begin{align}
      {\mathbb{E}} \left[ {\sum\limits_i {{Y_i}} } \right] &= \sum\limits_i {{k_i}} {\theta _i},\\
      \mathbb{E} \left[ {{{\left( {\sum\limits_i {{Y_i}} } \right)}^2}} \right] &= \sum\limits_i {{k_i}} \theta _i^2 + {\left( {\sum\limits_i {{k_i}} {\theta _i}} \right)^2},\\
      {\rm{Var}}\left[ {\sum\limits_i {{Y_i}} } \right] &= \sum\limits_i {{k_i}} \theta _i^2,  
  \end{align}
\end{subequations}

Therefore, the distribution of $\sum\limits_i {{Y_i}}$, which obey Gamma distribution can be approximated as
\begin{equation}
  \sum\limits_i {{Y_i}} \sim \Gamma \left( {\frac{{{{\left( {\sum\limits_i {{k_i}} {\theta _i}} \right)}^2}}}{{\sum\limits_i {{k_i}} \theta _i^2}},\frac{{\sum\limits_i {{k_i}} \theta _i^2}}{{\sum\limits_i {{k_i}} {\theta _i}}}} \right).
\end{equation}

Also, in (\ref{Schur_app}), we can obtain:
\begin{equation}
  \label{Schur_gamma}
  {\frac{{\det \left( {{{\bf{G}}_i}} \right)}}{{\det \left( {{{\bf{G}}_{i,k}}} \right)}}} \approx \sum\limits_{l \in {{\widetilde {\cal L}}_{m,k}}} {{{\left| {{h_{k,l}}} \right|}^2}}  = {\Lambda _{k,m}}\sim \Gamma ({\Psi _{k,m}},{\Phi _{k,m}}),
\end{equation}
where ${\Phi _{k,m}} \buildrel \Delta \over = \frac{{\sum\limits_{\tilde l \in {{\widetilde {\cal L}}_{m,k}}} {\beta _{\tilde l,k}^2} }}{{\sum\limits_{\tilde l \in {{\widetilde {\cal L}}_{m,k}}} {{\beta _{\tilde l,k}}} }}$,${\Psi _{k,m}} \buildrel \Delta \over = \frac{{{{\left( {\sum\limits_{\tilde l \in {{\widetilde {\cal L}}_{m,k}}} {{\beta _{\tilde l,k}}} } \right)}^2}}}{{\sum\limits_{\tilde l \in {{\widetilde {\cal L}}_{m,k}}} {\beta _{\tilde l,k}^2} }}$.

Therefore, due to the independence of channels and Gamma approximation lemma 1, the summation of multiple Gamma approximations still follows a Gamma distribution. The form of summation after taking expectations should theoretically be consistent with the direct approximation form. So from (\ref{Schur_gamma}) and (\ref{channel}), the upper bound of $X$ can be expressed as,
\begin{equation}
  \label{R_upper}
  \begin{aligned}
    {X^{{\rm{ub}}}} &= \sum\limits_{i = 1}^K {{{\log }_2}\left( {1 + {p_i}{\mathbb{E}_{{{\bf{H}}_1},{{\bf{H}}_2}, \ldots ,{{\bf{H}}_M}}}\left\{ {\sum\limits_{m = 1}^M {{\Lambda _{i,m}}} } \right\}} \right)}\\
    &={\log _2}\left( {1 + {p_i}\left\{ {\sum\limits_{m = 1}^M {\sum\limits_{\tilde l \in {{\widetilde {\cal L}}_{m,i}}} {{\beta _{\tilde l,i}}} } } \right\}} \right).
  \end{aligned}
\end{equation}

The upper bound of $Y$ can be expressed as,
\begin{equation}
  \label{V_upper}
  \begin{aligned}
    &{Y^{{\rm{ub}}}} \\
    &= \frac{{{Q}^{-1}}\left( {{\varepsilon}} \right)}{\sqrt{n}}\sqrt {K - \sum\limits_{i = 1}^K {1/{{\left( {1 + {p_i}\left\{ {\sum\limits_{m = 1}^M {\sum\limits_{\tilde l \in {{\widetilde {\cal L}}_{m,i}}} {{\beta _{\tilde l,i}}} } } \right\}} \right)}^2}}}.
  \end{aligned}
\end{equation}

On the other hand, the lower bound on SE of LHS and RHS in (\ref{uplink_Avg_ZF_SINR_lower}) also can be analyzed. Similar to \cite{liu2019spectral}, we introduce the Inverted Gamma distribution where ${X_{k,m}} = \frac{1}{{{\Lambda_{k,m}}}}{\sim }{\bf{Inverted}}{\;}{\bf{Gamma}}({\Psi_{k,m}},{\Phi_{k,m}})$,with ${\Phi_{k,m}} \buildrel \Delta \over = \frac{{\sum\limits{\tilde l \in {{\widetilde {\cal L}}_{m,k}}} {\beta_{\tilde l,k}^2} }}{{\sum\limits{\tilde l \in {{\widetilde {\cal L}}_{m,k}}} {{\beta_{\tilde l,k}}} }}$ and ${\Psi_{k,m}} \buildrel \Delta \over = \frac{{{{\left( {\sum\limits{\tilde l \in {{\widetilde {\cal L}}_{m,k}}} {{\beta_{\tilde l,k}}} } \right)}^2}}}{{\sum\limits{\tilde l \in {{\widetilde {\cal L}}_{m,k}}} {\beta_{\tilde l,k}^2} }}$. Moreover, the expectation of this random variable satisfies $\mathbb{E}\left\{{{X_{k,m}}} \right\} = \frac{1}{{{\Phi _{k,m}}\left( {{\Psi _{k,m}} - 1} \right)}}$.

Therefore, based on the above derivation, we can conclude,
\begin{equation}
  \label{Inverted_Gamma}
  \begin{aligned}
  \frac{1}{{\sum\limits_{m = 1}^M {{{\left\| {{{\bf{v}}_{k,m}}} \right\|}^2}} }} &= \frac{1}{{\sum\limits_{m = 1}^M {\frac{{\det \left( {{{\bf{Z}}_{m,k}}} \right)}}{{\det \left( {{{\bf{Z}}_m}} \right)}}} }}\\
  &\mathop  \approx \limits^{{L_m} \gg K} \frac{1}{\sum\limits_{m = 1}^M {\left( {\frac{1}{{\sum\limits_{l \in {{\widetilde {\cal L}}_{m,k}}} {{{\left| {{h_{k,l}}} \right|}^2}} }}} \right)}}= \frac{1}{{\sum\limits_{m = 1}^M {{X_{k,m}}}}}.
  \end{aligned}
\end{equation}

\textit{Remark 3: We employed a similar approach to Schur's Complementary Lemma and gamma approximation, as presented in \cite{liu2019spectral}, to handle the matrix ${{{\left[ {{{\left( {{\bf{H}}_m^{\rm{H}}{{\bf{H}}_m}} \right)}^{ - 1}}} \right]}_{kk}}}$ in our combining vector corresponding to the EDUs. By exploiting the independence of the channel for each EDU, we expressed the resulting SINR in a summation form, as shown in (\ref{inverse_formular}) and (\ref{Inverted_Gamma}). 
For the upper bound result (\ref{uplink_Avg_ZF_SINR_upper}), further manipulation of (\ref{inverse_formular}) is required to obtain the final expression by Jensen's inequality (\ref{inverse_formular_jensen}).}

Therefore, the lower bound of LHS $X$ (\ref{uplink_Avg_ZF_SINR_lower}) can be expressed as:

\begin{equation}
  \label{R_lower}
  \begin{aligned}
    X \ge {X^{{\rm{lb}}}} &= \sum\limits_{i = 1}^K {{{\log }_2}\left( {1 + \frac{{{p_i}{M^2}}}{{{\mathbb{E}_{{{\bf{H}}_1},{{\bf{H}}_2}, \ldots ,{{\bf{H}}_M}}}\left\{ {\sum\limits_{m = 1}^M {{X_{k,m}}} } \right\}}}} \right)}\\
    & = \sum\limits_{i = 1}^K {{{\log }_2}\left( {1 + \frac{{{p_i}{M^2}}}{{\sum\limits_{m = 1}^M {\frac{1}{{{\Phi _{i,m}}\left( {{\Psi _{i,m}} - 1} \right)}}} }}} \right)}.
  \end{aligned}
\end{equation}

Similarly, the lower bound of RHS $Y$ can be expressed as,

\begin{equation}
  \label{V_lower}
  \begin{aligned}
    &Y \ge {Y^{{\rm{lb}}}} = \\
    &= \frac{{{Q}^{-1}}\left( {{\varepsilon}} \right)}{\sqrt{n}}\sqrt {K - \sum\limits_{i = 1}^K {1/{{\left( {1 + \frac{{{p_i}{M^2}}}{{\sum\limits_{m = 1}^M {\frac{1}{{{\Phi _{i,m}}\left( {{\Psi _{i,m}} - 1} \right)}}} }}} \right)}^2}}}. 
  \end{aligned}
\end{equation}

Therefore, considering finite block length, for the sum SE of the system, we have:

\begin{equation}
  \label{ALL_bound}
  {R^{{\rm{lb}}}} = {X^{{\rm{lb}}}} - {Y^{{\rm{ub}}}}  \le \bar R \le {X^{{\rm{ub}}}} - {Y^{{\rm{lb}}}}  = {R^{{\rm{ub}}}}.
\end{equation}

In the following sections, we will validate the effectiveness of the boundaries.

Clearly, for the scenario of a single EDU, (\ref{R_upper}), (\ref{R_lower}) align with the form in \cite{liu2019spectral}. When all RRUs are located at the same position, their upper and lower bound results align with the form of traditional massive MIMO in the uRLLC case.

\begin{figure}[!h]
  \centering
  \includegraphics[width=3.5in]{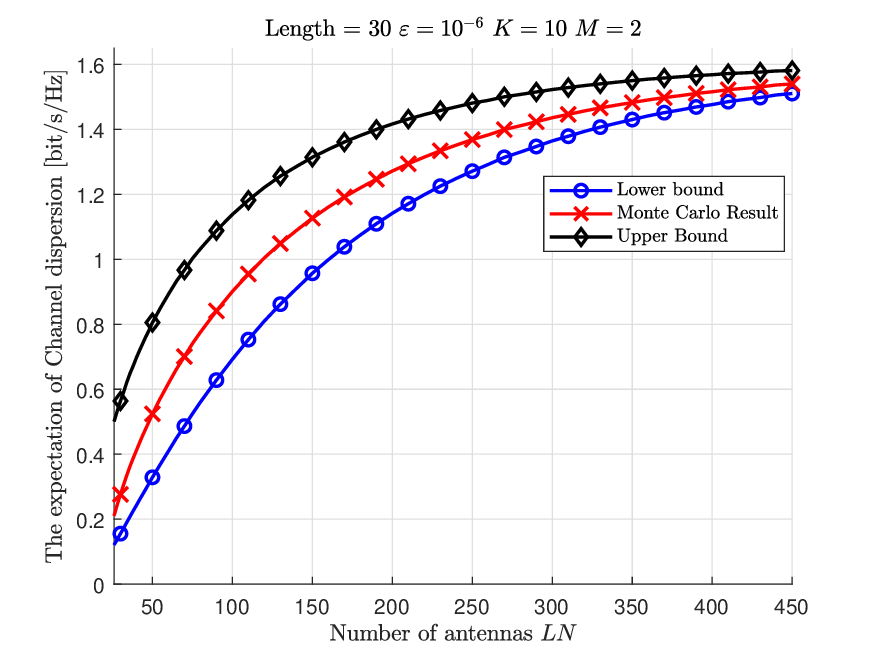}
  \caption{Channel dispersion varies with the number of receiving antennas, $M=2$,$K=10$,$n=50$,$\varepsilon=10^{-6}$.}
  \label{fig2_Vcdf}
\end{figure}

In Fig.\ref{fig2_Vcdf}, we verify the upper and lower bounds derived for the channel dispersion term $Y$ in (\ref{V_lower}) and (\ref{V_upper}). We simulated the expected performance of the channel dispersion $V$ for a scenario with block length $n=30$ and transmission error probability $\varepsilon = 10^{-6}$, considering UE and EDU number are $K=10$ and $M=2$, respectively.

Simulation results indicate that our derived bounds provide tighter gap limits than Monte Carlo simulations. As the number of receiving antennas $LN$ increases, the system's SINR significantly improves. It is also observed that the computation of $Y$ stabilizes when the number of antennas exceeds $60$.

Due to the scalability of cell-free systems where the number of receiving antennas is far more than that of transmitting antennas, even in scenarios with high transmission error probability requirements and short block lengths, the expected value of channel dispersion $Y$ remains significantly lower than the system's rate. Therefore, our subsequent analysis on SE under finite block length, focusing on the first part of (\ref{uplink_sum_Rate}), specifically the traditional channel capacity component, will be optimized.

\subsection{Large-scale Analysis}

For analytical convenience and without loss of generality, we adopt the free space path loss model so the large scale channel model is given by
\begin{equation}
  \label{largr_fading}
  \beta_{k,l} = d_{k,l}^{-\alpha},
\end{equation}
where $d_{k,l}$ represents the distance from the UE $k$ to RRU $l$ and $\alpha$ denotes the path loss exponent and we note that $d_{k,l} > 0$ holds under all circumstances.

We aim to maximize UE $k$'s SE for the large-scale fading coefficient. This problem is formulated using its SE upper or lower bound (\ref{R_upper}), (\ref{R_lower}). We use the upper bound to transform the problem into,

\begin{equation}
  \label{max_question}
  \max\left\{ {\log _2}\left( {1 + {p_k}\left\{ {\sum\limits_{m = 1}^M {\sum\limits_{\tilde l \in {{\widetilde {\cal L}}_{m,k}}} {{\beta _{\tilde l,k}}} } } \right\}} \right) \right\}.
\end{equation}

Optimizing the form of (\ref{max_question}) is challenging due to its complexity. Consider the function $f(x)=x^{-\alpha}$, which is convex because $f''(x)=\alpha(\alpha+1)x^{-\alpha-2}>0$ holds for all $x>0$. Therefore, we can apply Jensen's inequality:

\begin{equation}
  \label{distance_jensen}
  \frac{1}{n} \sum_{i=1}^n d_i{ }^{-\alpha}=\frac{1}{n} \sum_{i=1}^n f\left(d_i\right) \geq f\left(\frac{1}{n} \sum_{i=1}^n d_i\right)=\left(\frac{1}{n} \sum_{i=1}^n d_i\right)^{-\alpha}.
\end{equation}

To facilitate further processing, we represent the above expression in terms of squared distance,
\begin{equation}
  \label{distance_jensen_power2}
  \begin{aligned}
    \frac{1}{n}{\sum\limits_{i = 1}^n {{d_i}} ^{ - \alpha }}& = \frac{1}{n}{\sum\limits_{i = 1}^n {\left( {{d_i}^2} \right)} ^{ - \frac{\alpha }{2}}} = \frac{1}{n}\sum\limits_{i = 1}^n f \left( {{d_i}^2} \right)\\
    & \ge f\left( {\frac{1}{n}\sum\limits_{i = 1}^n {{d_i}^2} } \right) = {\left( {\frac{1}{n}\sum\limits_{i = 1}^n {{d_i}^2} } \right)^{ - \frac{\alpha }{2}}}.
  \end{aligned}
\end{equation}

According to (\ref{distance_jensen_power2}), we can approximate the (\ref{max_question}) problem and obtain its lower bound as follows,
\begin{equation}
  \label{max_question_lower}
  R_k^{{\rm{ub}}}{\rm{ }} \ge {\log _2}\left( {1 + {p_k}\left\{ {\sum\limits_{m = 1}^M {{{\left( {\sum\limits_{\tilde l \in {{\widetilde {\cal L}}_{m,k}}} {{d_{k,l}}}^2 } \right)}^{ - \frac{\alpha }{2}}}} } \right\}} \right).
\end{equation}

By maximizing its lower bound, the original problem of maximizing SINR is transformed into a problem of minimizing the sum of distances from UE $k$ to all RRUs formulated as follows,
\begin{equation}
  \label{max_question_change}
  {\rm{max}}\left\{ {\sum\limits_{m = 1}^M {\sum\limits_{\tilde l \in {{\widetilde {\cal L}}_{m,k}}} {{\beta _{\tilde l,k}}} } } \right\} \to {\rm{min}}\left\{ {\sum\limits_{m = 1}^M {{{\left( {\sum\limits_{\tilde l \in {{\widetilde {\cal L}}_{m,k}}} {{{d_{k,l}}}^2} } \right)}}} } \right\}.
\end{equation}

Assuming there are $L$ RRUs, the coordinates of these RRUs are $(x_1, y_1), (x_2, y_2), \cdots, (x_L, y_L)$ and the coordinates of UE $k$ are $(x_k, y_k)$. The distance from UE $k$ to RRU $l$ is represented as follows,
\begin{equation}
  \label{distance_two_point}
  d_{k,l}=\sqrt{\left(x_k-x_l\right)^2+\left(y_k-y_l\right)^2}.
\end{equation}

The distance and square of UE $k$ to all RRUs are represented as follows,
\begin{equation}
  \label{distance_all_point}
  d_k^2=\left( {\sum\limits_{\tilde l \in {{\widetilde {\cal L}}_{m,k}}} {{d_{k,l}}}^2 } \right)={\sum\limits_{\tilde l \in {{\widetilde {\cal L}}_{m,k}}} \left[\left(x_k-x_l\right)^2+\left(y_k-y_l\right)\right]^2}.
\end{equation}

Let us consider the presence of an EDU in the scenario, representing traditional centralized processing. In order to minimize $d_k^2$, we assume that the UEs follow a uniform distribution; therefore, by applying the concavity and convexity of the associative function, taking partial derivatives concerning $x$ and $y$ for the sum of squared distances (\ref{distance_all_point}) yields the following expression,
\begin{equation}
  \begin{aligned}
    & \frac{\partial d_k^2}{\partial x_k}=2 \sum\limits_{\tilde l \in {{\widetilde {\cal L}}_{m,k}}}\left(x_k-x_l\right), \\
    & \frac{\partial d_k^2}{\partial y_k}=2 \sum\limits_{\tilde l \in {{\widetilde {\cal L}}_{m,k}}}\left(y_k-y_l\right).
  \end{aligned}
\end{equation}

Based on the necessary conditions for extreme values, we equate the partial derivatives of $d_k^2$ concerning $x_k$ and $y_k$ to $0$ as,
\begin{equation}
  \begin{aligned}
    & x_k=\frac{1}{n} \sum\limits_{\tilde l \in {{\widetilde {\cal L}}_{m,k}}} x_l, \\
    & y_k=\frac{1}{n} \sum\limits_{\tilde l \in {{\widetilde {\cal L}}_{m,k}}} y_l.
  \end{aligned}
\end{equation}

That is, the distance from UE $k$ to all RRUs is the smallest, and only if UE $k$ is located at the center of gravity of all RRUs is the sum of distance (\ref{distance_all_point}) the smallest; that is, the lower bound (\ref{max_question_change}) is the largest.

Based on equation (\ref{max_question_lower}), we consider the scenario involving multiple EDUs. In this paper, the UEs should be located at the center of gravity of the RRUs in multiple groups of EDUs, while the center of gravity of the UEs should be positioned at the midpoint of the plane. Consequently, each group of RRUs can exhibit a uniform distribution to achieve optimal performance.

However, as the number of EDUs increases, satisfying this requirement becomes more challenging. As a result, achieving a statistically uniform distribution in a higher-dimensional space becomes increasingly tricky. To address this issue, in section IV, we propose a improved graph coloring algorithm that aims to interleave the RRUs between groups as much as possible, considering the random distribution of RRUs. The superiority of the proposed interleaving deployment method has been verified in actual systems, especially in scenarios with OTA reciprocity calibration \cite{cao2023experimental}.

\subsection{RRU correlation method}

A genetic algorithm based on the distance relationship between RRUs was proposed to associate EDUs and RRUs \cite{cao2023oran}. The designed fitting function was,
\begin{equation}
  \label{GA_fitness}
  \begin{aligned}
    f(x) = \frac{1}{\sum\limits_{p\in \mathcal{P}}{\sum\limits_{q\in \mathcal{Q}}{\cdots }}\sum\limits_{u\in \mathcal{U}}{\sum\limits_{v\in \mathcal{V}}{\left( {{d}_{p,q}^{2}}+\cdots +{{d}_{p,v}^{2}}+\cdots +{{d}_{u,v}^{2}} \right)^{1/2}}}},    
  \end{aligned}
\end{equation}
where ${d}_{p,v}$ is the distance between the $p-$th RRU and the $v-$th RRU, $\mathcal{P},\mathcal{Q},\cdots,\mathcal{U},\mathcal{V}$ is the $p,q, \cdots,u,v$ 's corresponding RRU grouping of EDUs, and $\mathcal{T}$ is a complete set of all RRUs.

The objective function of the heuristic scheme becomes highly complex with the increase in the number of EDUs, and this heuristic Algorithm has not been theoretically analyzed.
Therefore, based on our theoretical analysis above, Assuming the UE's centroid is located at the origin, we use a graph coloring algorithm to reduce the implementation complexity. This scheme's complexity is significantly reduced compared to the method in the original paper.

\subsection{Improved Graph Coloring Algorithm}

Based on the analysis in the previous section, we know that for multiple groups of EDUs, the RRUs under each group of EDUs should align as closely as possible with the geometric centroid of the UE locations. Since UEs are randomly distributed, we also need to ensure that multiple groups of EDUs satisfy an interleaved random distribution. Therefore, to avoid concentrating RRUs in a small area, we use an improved graph coloring algorithm for association.

In scenarios with a large number of RRUs, especially when the number of RRUs exceeds $100$, the chromosome population designed by traditional genetic algorithms will be huge, leading to enormous complexity \cite{cao2023oran}. This is an NP-hard combinatorial optimization problem, but the graph coloring algorithm can effectively reduce the implementation complexity of the Algorithm \cite{liu2020graph}. The graph coloring process indicates a conflict if there is a connection and the two endpoints of the line need to use different coloring schemes, while unconnected RRUs can use the same color. To make the distribution as uniform as possible, we need to fully utilize the number of allocated colors, i.e., the number of EDUs in our case. Previous work has applied graph coloring algorithms to UE pilot allocation strategies \cite{zhu2015graph}. Here, we apply an improved graph coloring algorithm to the interleaved deployment of RRUs and EDUs.
\begin{figure}[!h]
  \subfloat[Step N-1:$\delta=0.125$,$M=5$.]{\includegraphics[width=3.5in]{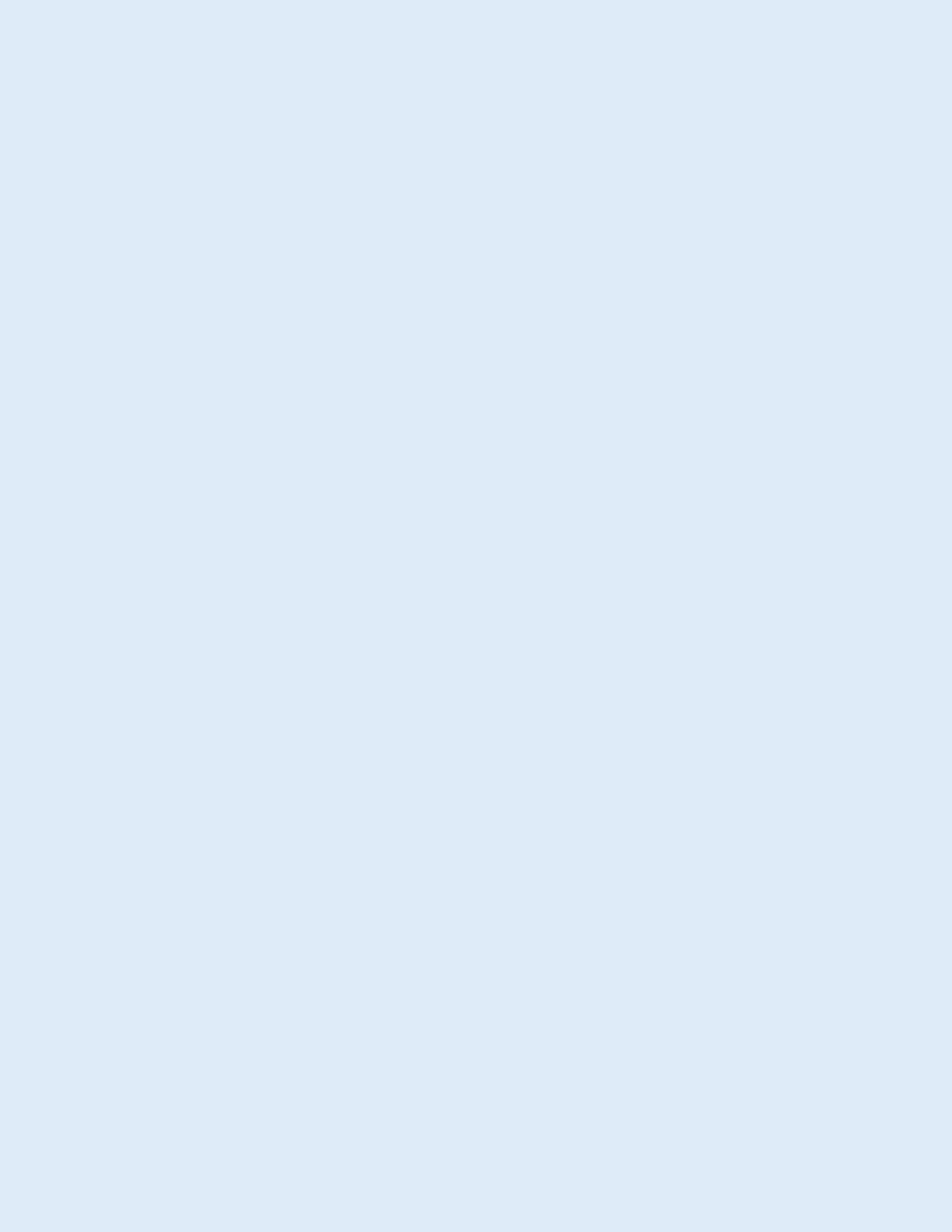}
  \label{stepNsub1}}\\
  \subfloat[Step N:$\delta=0.0625$,$M=2$.]{\includegraphics[width=3.5in]{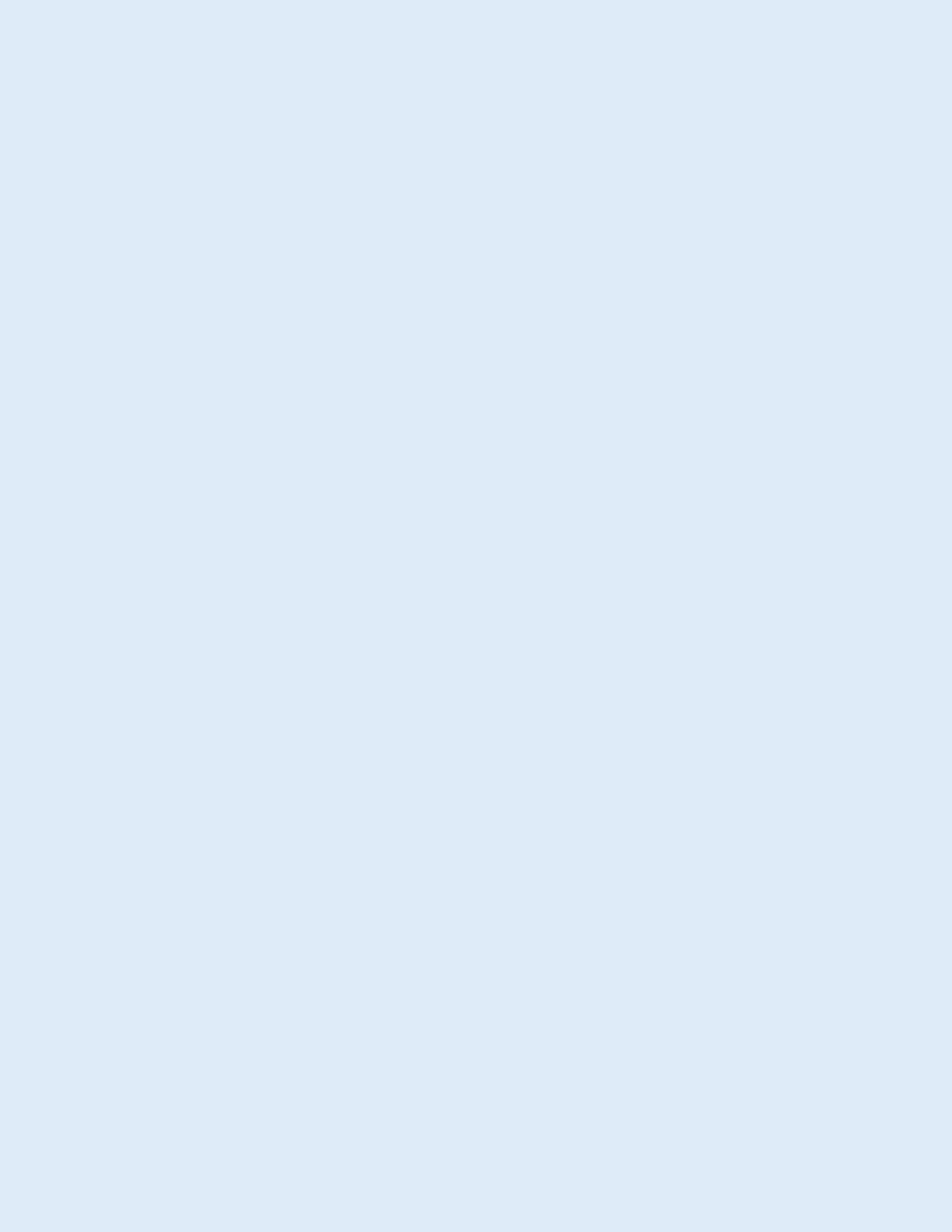}
  \label{stepN}}
  \caption{The graph coloring algorithm implementation process, $L=32$,$K=10$.}
  \label{fig4_Graphcolorstep}
\end{figure}

In scenarios with a large number of RRUs, there will inevitably be cases of unallocated conflicts. Therefore, based on the original graph coloring, we sort the unallocated conflicting RRUs and isolated RRUs according to their distances and connect them sequentially based on the number of remaining colors and distances, solving the problem of isolated points to ensure uniform distribution as much as possible.

The execution process of the improved graph coloring algorithm is shown in Algorithm \ref{alg3}, where the associated distance $\delta$ is continuously updated using the bisection method until the required number of colors meets the number of EDUs.

We use the tabu search method \cite{marappan2018solution} in the color allocation. Tabu search can gradually move towards the minimum value of the function. To avoid cycles and local minima, the tabu list is updated during iteration, reducing the search space and achieving rapid random convergence. The tabu search operation quickly searches for better approximate optimal solutions in a vast solution space.

We connect RRUs of different colors to the same EDU. The example in Fig.\ref{fig4_Graphcolorstep} shows the coloring process of $32$ RRUs with target $2$ EDUs. In the penultimate bisection search, five colors and the connection of RRUs to five EDUs are still required, necessitating further distance refinement. In the next bisection search, only two colors remain for the RRUs, meeting the requirements for the number of EDUs. The RRUs allocated by the improved graph coloring algorithm meet the conditions for maximum interleaving, and calculations are only needed during the initial deployment. Since the centroid positions of all RRUs are located at the system's origin, it supports infinite scalability. The system's performance will also improve with the increase in the number of EDUs and the deployment of RRUs.

\begin{algorithm}[h!]
  \caption{improved Graph Coloring Algorithm}
  \label{alg3}
  \begin{algorithmic}[1]
    \Statex \textbf{Input: }the number of RRU $L$, the number of EDU $M$, distance matrix $\textbf{D}\in \mathbb{R}^{L\times L}$, initial distance scale factor $\delta$.
    \Statex \textbf{Output: }RRU Group $\mathcal{L}_m$
    \State Every EDU chooses $L_m$ RRUs to generate the connect set $\mathcal{L}_m$.
    \State Generate associate graph matrix $\textbf{D}\in \mathbb{R}^{L\times L}$ through $\mathcal{L}_m$.
    \State According to $\textbf{D}$, use the graph coloring algorithm to generate graph results and calculate the target EDU number $n$ based on $\delta$.
    \State \textbf{while} $n\neq M$ \textbf{do}
    \State \qquad Use the bisection method to update $\delta$. 
    \State \qquad According to $\delta$, sort the positions by distance based on isolated points and RRUs that are not fully colored and connect to all colors that are fully colored.
    \State \qquad Use the graph coloring algorithm to recalculate $n$ and update.
    \State \textbf{end while}
 \end{algorithmic}
\end{algorithm}

\begin{table}[ht!]
  \small
  \caption{Simulation parameters} 
  \centering 
  \begin{threeparttable}[b]
  \begin{tabular}{l l}
  \toprule[1.5pt]
  \textbf{Simulation Parameters} &  \textbf{Values} \\
  \midrule
  Uplink transmission power                     &  $200\;\rm{mW}$ \\
  Antenna height                                &  10\;m \\
  Area size                                     &  $200\times 200\;\rm{m}^2$ \\
  The number of RRUs, $L$                       &  100 \\
  Number of antennas per RRU, $N$               &  4 \\
  Number of UEs, $K$                            &  24\\
  Number of orthogonal pilots,${L_P}$           &  24 \\
  Transmission bandwidth, $B$                   &  20\;MHz \\
  Carrier frequency, $f_c$                      &  2\;GHz \\
  Azimuth angle, $\bar{\varphi}$ (in radians)   &  15\;\\
  Elevation angle, $\bar{\theta}$ (in radians)  &  15 \\
  Noise power spectral density, $N_0$           &  $-174\;\rm{dBm/Hz}$ \\
  \midrule
  \bottomrule[1.5pt]
  \end{tabular}
  \end{threeparttable}
  \label{table_1}
\end{table}

\section{Simulation Results Analysis}

Based on the analysis in the previous sections, we conduct simulation verification. Firstly, we validate the proposed upper and lower bounds (\ref{ALL_bound}) against Monte-Carlo simulation results.
Subsequently, we further consider more realistic scenarios including imperfect CSI and pilot contamination, and compare the performance of different UE association strategies.

\subsection{Validation of Upper and Lower Bounds of Average SE in EDU Scenarios}

\begin{figure}[h!]
  \centering
  \subfloat[System uplink SE CDF verification with $n=50$, $\varepsilon=10^{-6}$, $K=10$, $L=300$, $N=1$.]{\includegraphics[width=3.5in]{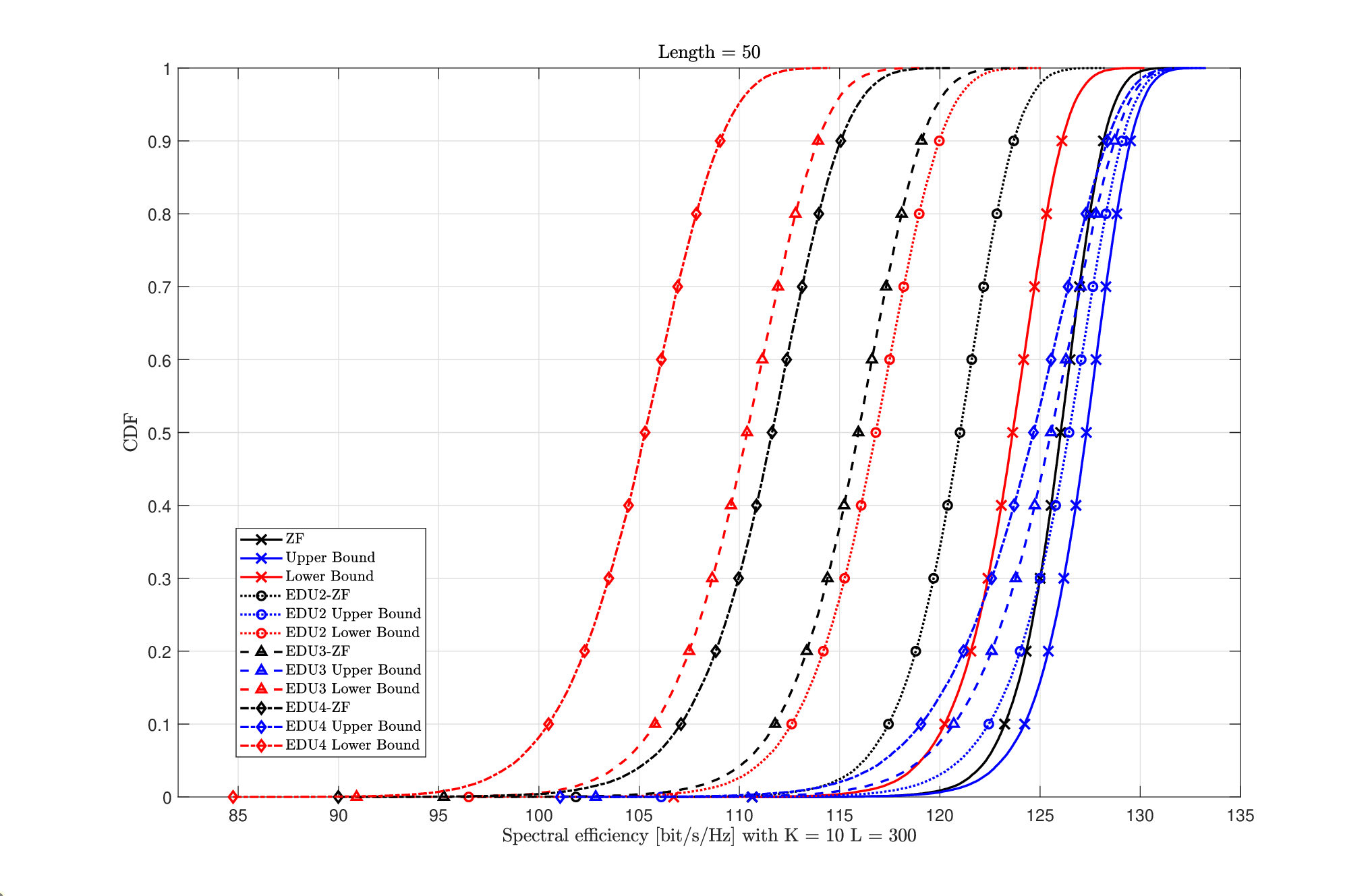}
  \label{fig3_a_cdf10}}\\
  \subfloat[System uplinkSE CDF verification with $n=50$, $\varepsilon=10^{-6}$, $K=30$, $L=300$, $N=1$.]{\includegraphics[width=3.5in]{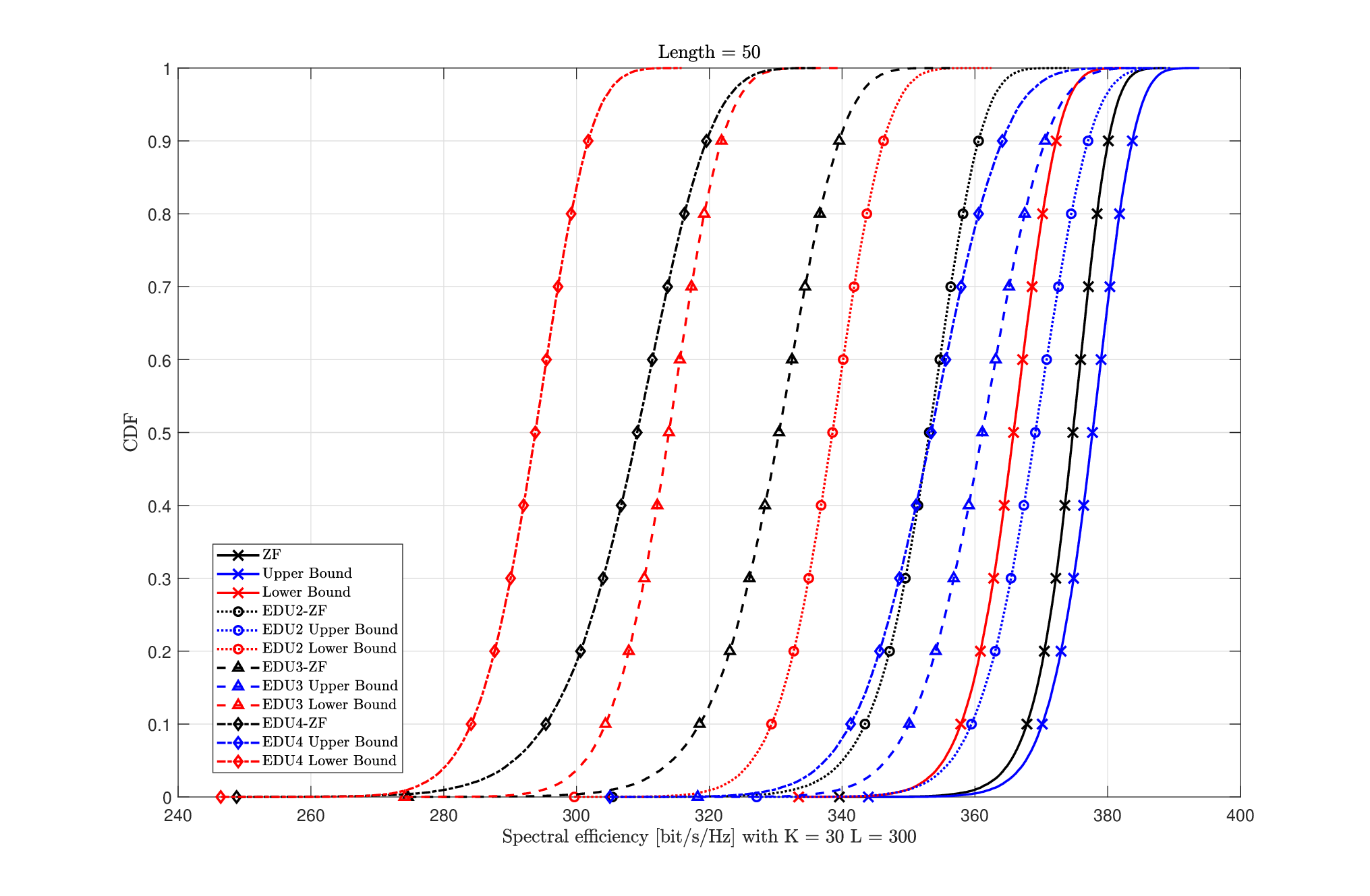}
  \label{fig3_b_cdf30}}
  \caption{System uplink SE CDF verification.}
  \label{fig3_CDFbound}
\end{figure}

In Fig.\ref{fig3_CDFbound}, we validate the perfect CSI by comparing the Monte Carlo simulation results using ZF combining with the theoretical results in (\ref{ALL_bound}).
To align as closely as possible with centralized processing in a cell-free system, we refer to the simulation settings in \cite{liu2019spectral}. The system consists of $L=300$ single-antenna RRUs and $K$ UEs uniformly distributed within a specific range.
We select a path loss factor of $\alpha =4$.
Fig.\ref{fig3_a_cdf10} and Fig.\ref{fig3_b_cdf30} simulate the performance for $K=10$ and $K=30$ UEs, respectively.
From the gap between the grouped EDUs and theoretical values, we can see that the system performance deteriorates as the number of EDU groups increases.
It can be seen that as the number of UEs increases, the system's SE continuously improves, and under different numbers of EDUs, the upper and lower bounds of the SE with finite block length can maintain good approximation.
This result verifies the validity of the derived performance upper and lower bounds in (\ref{ALL_bound}).

\subsection{Simulation Performance of Interleaved EDU Grouping}

In this subsection, we consider more complex scenarios for simulation. We account for imperfect CSI in the uplink system, with large-scale fading as follows \cite{bjornson2017massive},
\begin{equation}
  \beta_{k,l}(\textrm{dB}) = -30.5 - 36.7 \log_{10}\left( {d_{k,l}} \right)  + F_{k,l},
\end{equation}
where $F_{k,l}$ represents the impact of shadow fading.

Due to the noise power amplification issue in ZF combining, as shown in (\ref{uplink_ZF_SINR}), we employ MMSE channel combining scheme and consider the impact of multiple antennas, assuming $N=4$. The simulation parameters are shown in Table \ref{table_1}.
Due to channel estimation errors and the limited number of pilots, imperfect CSI will be introduced.
Considering pilot contamination, we further associate UE and RRU and the association information can be obtained at the EDU.
We use the classic scalable association method, dynamic cooperation clusters (DCC), to deal with the pilot contamination in a cell-free scenario \cite{bjornson2020scalable}.
So, the combining vector at the EDU considers the influence of the UE-RRU association matrix, and we adopt the MMSE combining method with the corresponding vector as follows,
\begin{equation}
  \label{MMSE}
  \begin{aligned}
    &{\bf{v}} _{k,m} = \\
    &{p_k}{\left( {\sum\limits_{i = 1}^K {{p_i}{{\bf{D}}_{k,m}}\left( {{{\widehat {\bf{h}}}_{i,m}}\widehat {\bf{h}}_{i,m}^{\rm{H}}} \right){{\bf{D}}_{k,m}}}  + \sigma _{{\rm{UL}}}^2{{\bf{I}}_{mN}}} \right)^{ - 1}}{{\bf{D}}_{k,m}}{\widehat {\bf{h}}_{k,m}},
  \end{aligned}
\end{equation}
where ${\bf{D}}_{k,m}$ represents the association between the $m$-th EDU and UE $k$, ${{\hat{\mathbf{h}}}_{i,k}}$,${{\tilde{\mathbf{h}}}_{i,k}}$ represent the channel estimation information and channel estimation error respectively and ${{\mathbf{h}}_{i,k}}={{\hat{\mathbf{h}}}_{i,k}}+{{\tilde{\mathbf{h}}}_{i,k}}$.

We will compare the interleaving deployment strategy with traditional clustering grouping. The clustering strategy selects the initial position of the RRU as the starting centroid, with the number of EDUs $M$ as the number of cluster centroids, using the final clustering strategy of the RRUs as the corresponding association method. Considering that the centroid positions of the clusters will continuously change during the algorithm implementation, we randomly set the initial cluster centroids and use the final centroids as the deployment locations of the EDUs to minimize clustering error. This approach is common in traditional cellular architectures and CoMP systems, and in this collaborative architecture, we still use clustering as our comparison method. We adopt the classic K-means++ clustering strategy \cite{arthur2006k}, which is used in many scenarios due to its low complexity, fast convergence speed, and good performance \cite{le2021learning}.

\begin{figure}[!h]
  \centering
  \subfloat[$K=24$,without pilot contamination.]{\includegraphics[width=3.5in]{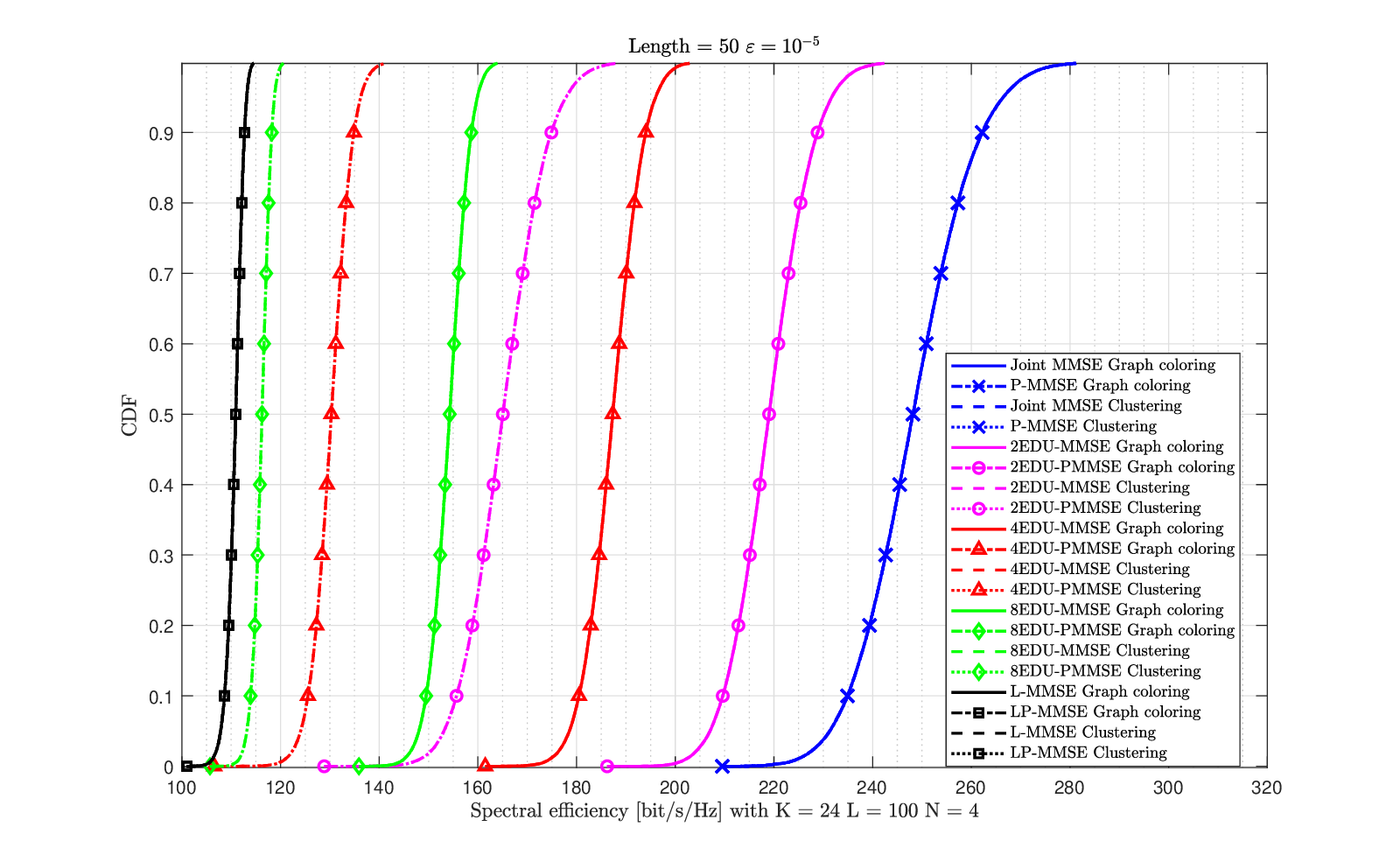}
  \label{fig5_a_secdfk24}}\\
  \subfloat[$K=48$,with pilot contamination.]{\includegraphics[width=3.5in]{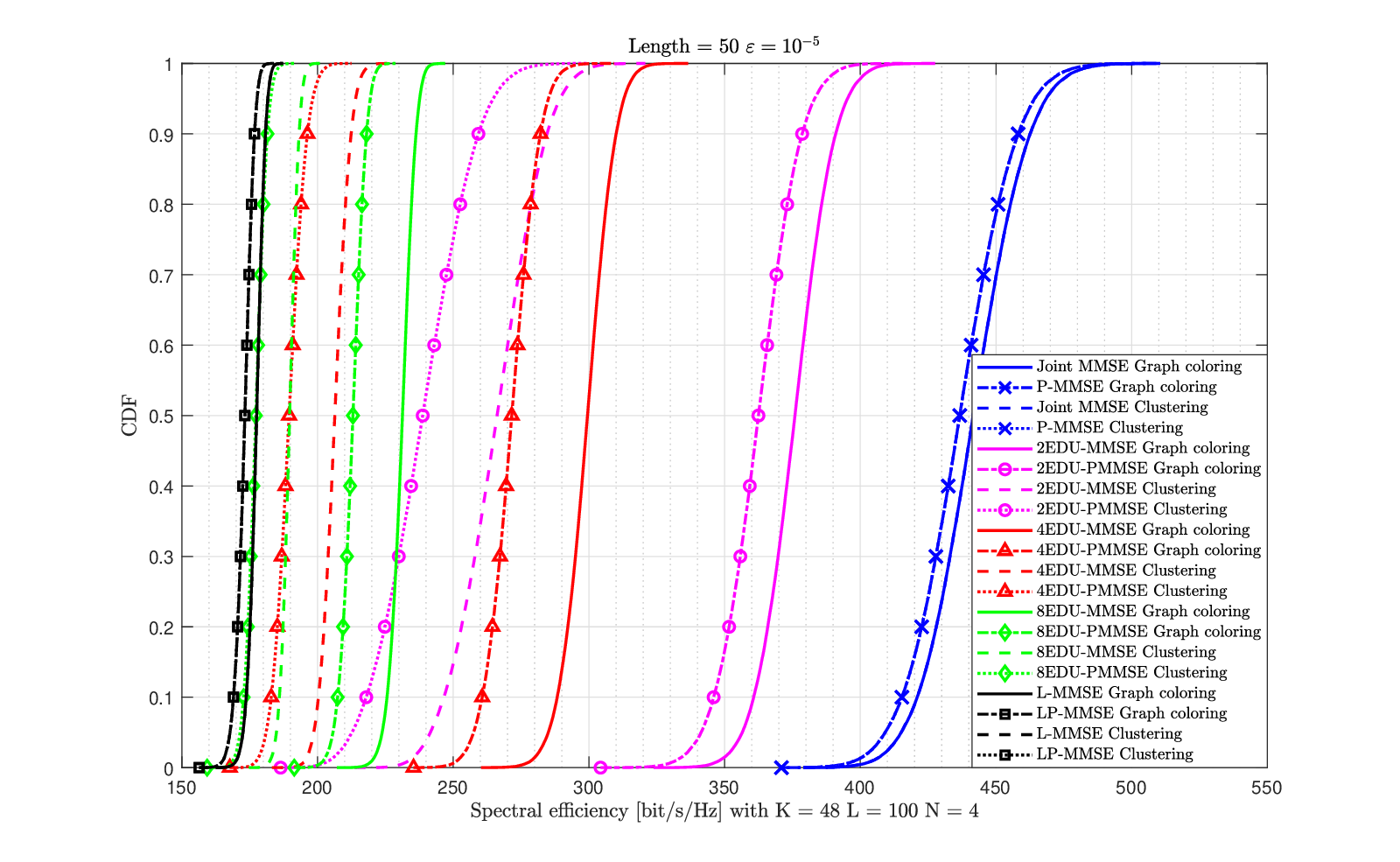}
  \label{fig5_b_secdfk48}}
  \caption{Comparison of uplink SE between clustering and improved graph coloring strategies under different numbers of EDUs with $n=50$, $\varepsilon=10^{-6}$, $L=300$, $N=1$.}
  \label{fig5_secdf}
\end{figure}

Fig.\ref{fig5_secdf} simulates the combining strategy shown in (\ref{MMSE}), which we refer to as EDU-MMSE and EDU-P-MMSE. Joint MMSE and LP-MMSE (local partial-MMSE) are both exceptional cases.
We further compare the performance of different EDU and RRU association strategies under the scenario of uplink transmission with or without pilot contamination. We simulate the scenario with transmission error probability of $\varepsilon=10^{-5}$ and a block length of $n=50$.
When the number of UE $K$ is greater than the number of pilots $L_P$, the DCC strategy is adopted for scalable EDU deployment.
Fully centralized processing has the highest implementation complexity and optimal performance, while fully distributed processing has the lowest implementation complexity and the worst performance.

As can be seen from Fig.\ref{fig5_a_secdfk24}, in the absence of pilot contamination, the SE with finite block length is the same when using the DCC strategy and fully centralized deployment, which means that UEs can all be allocated orthogonal pilots. In Fig.\ref{fig5_b_secdfk48}, the user-centric DCC strategy reduces centralized processing complexity at the cost of some performance loss.
Comparing the performance of the clustering strategy based on the K-means++ method and the interleaving strategy deployment based on the improved graph coloring algorithm, it can be seen that whether or not there is pilot contamination, the SE performance of the interleaving strategy is far superior to that of the clustering strategy under both fully centralized and partial cooperation processing. As the number of EDUs increases, the gap between the clustering strategy and the improved graph coloring method gradually decreases, indicating that the interleaving strategy cannot meet the absolute overlap of the center as analyzed in section III.
\begin{figure}[!h]
  \centering
  \subfloat[$K=24$,without pilot contamination.]{\includegraphics[width=3.5in]{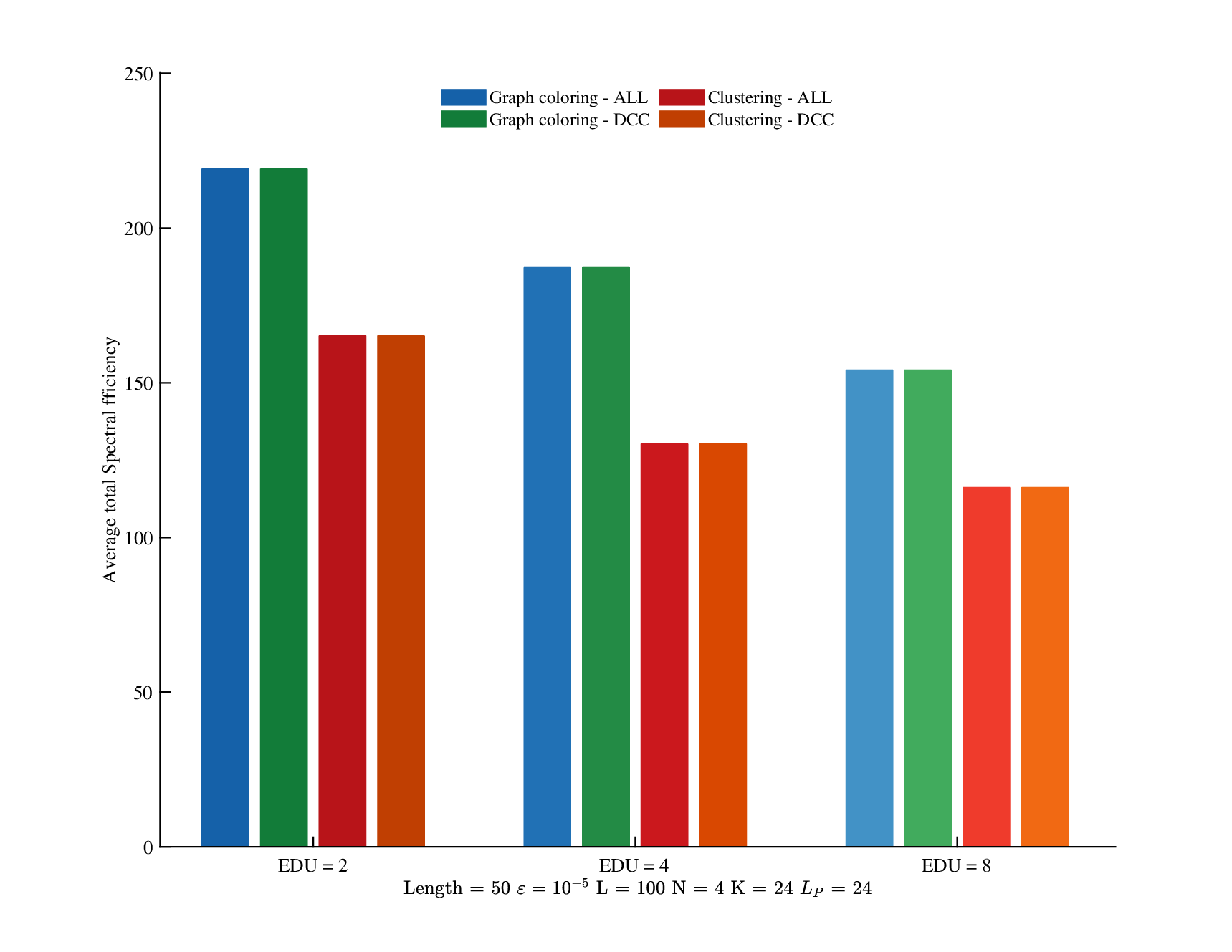}
  \label{fig6_a_Ese24}}\\
  \subfloat[$K=48$,with pilot contamination.]{\includegraphics[width=3.5in]{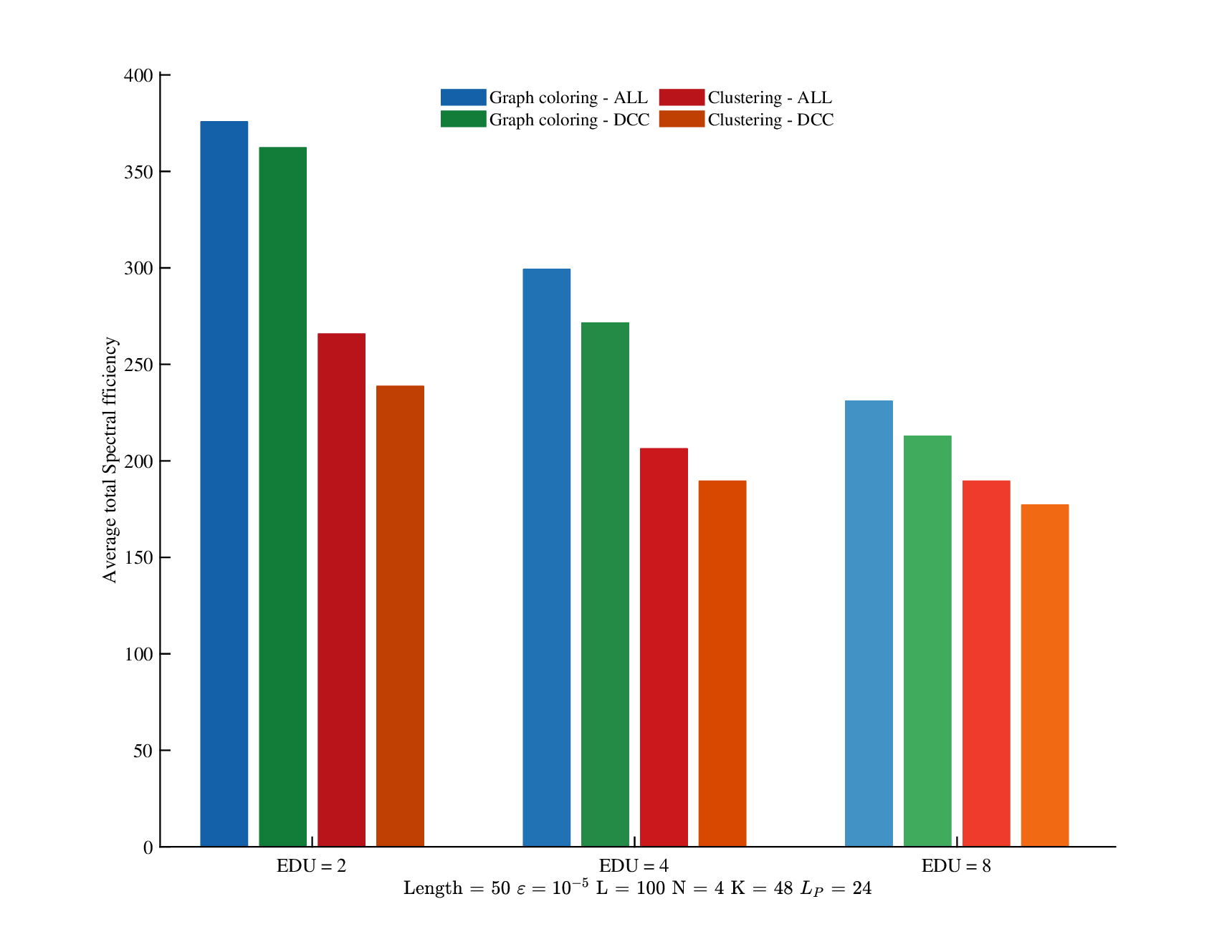}
  \label{fig6_b_Ese48}}
  \caption{Comparison of average total SE between clustering and improved graph coloring strategies under different numbers of EDUs with $n=50$, $\varepsilon=10^{-6}$, $L=300$, $N=1$.}
  \label{fig6_Ese_compare}
\end{figure}

In Fig.\ref{fig6_Ese_compare}, we compare the expected SE of finite block length under scenarios with the number of EDUs being $M=2$, $M=4$, and $M=8$. When the number of EDUs is $M=2$, the SE of finite block length using interleaving deployment can be increased by more than 30\% compared to the K-means++ clustering deployment.
The deployment method based on the improved graph coloring algorithm interleaving will be used unless otherwise specified In following simulations.

\textit{Remark 4: The interleaving deployment strategy aligns with the system's actual architectural deployment. This means that in a hotspot area, we can enhance system performance by continuously increasing the deployment density of RRUs and EDUs without considering the regional division of traditional cellular RRU clustering deployment.}

\begin{figure}[ht!]
  \centering
  \subfloat[$M=4$.]{\includegraphics[height=2.5in]{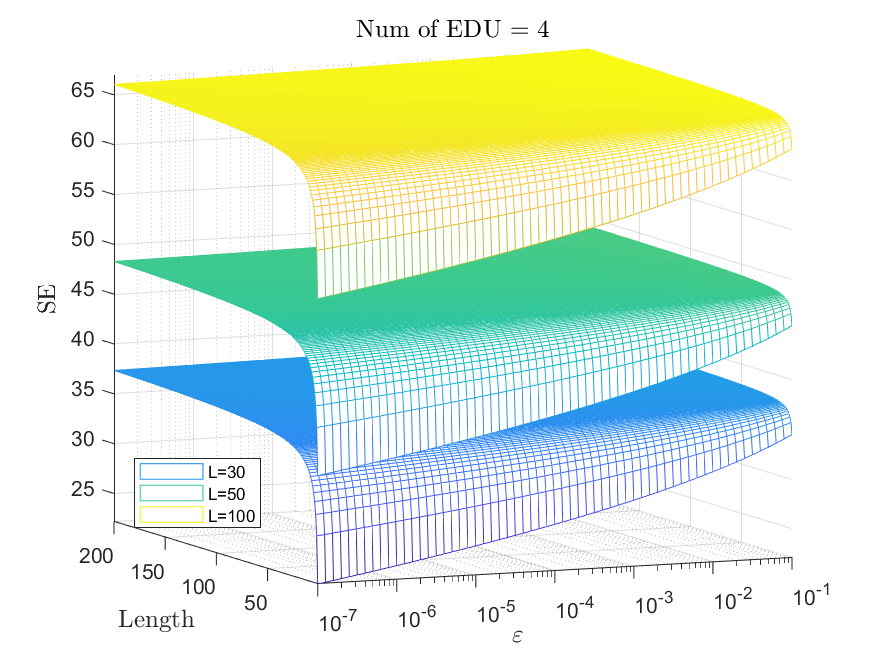}
  \label{fig7_a_3D_M4}}\\
  \subfloat[$M=2$.]{\includegraphics[height=2.5in]{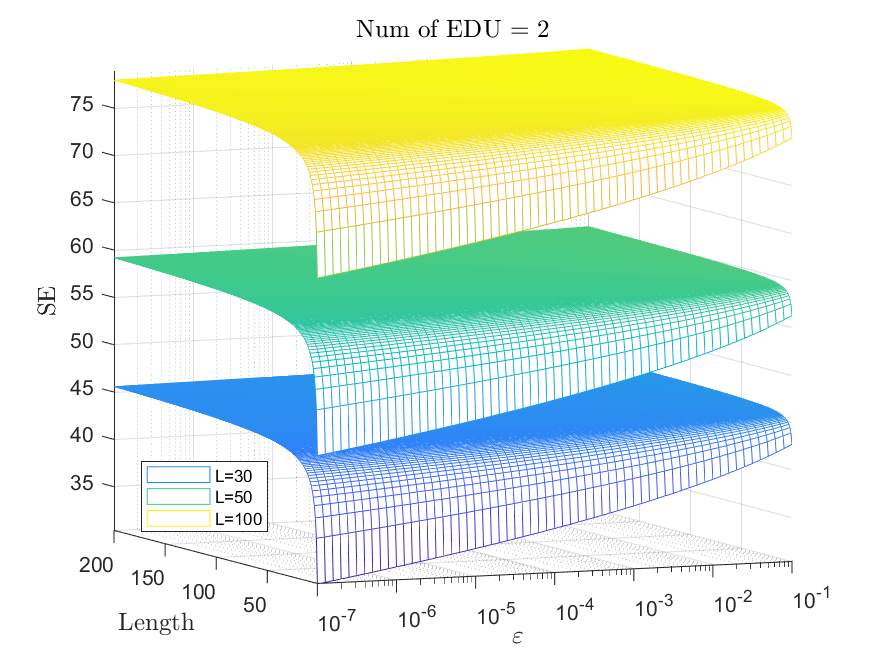}
  \label{fig7_b_3D_M2}}
  \caption{The relationship between system SE and transmission error probability $\varepsilon$ and block length $n$ varies with different numbers of RRUs.}
  \label{fig7_3D_compare}
\end{figure}

In Fig.\ref{fig7_3D_compare}, we simulate the relationship between the SE of finite block lengths and the transmission error probability under different numbers of RRUs. The capacity collapse effect persists regardless of the number of antennas. This is because, in finite block length scenarios, the size of channel dispersion $V$ becomes non-negligible as the block length decreases sharply. However, regardless of the codeword length $n$, the SE of finite block length will increase with the number of system antennas, as proposed by the spatiotemporal exchangeability in the \cite{you2023closed}. As shown in the comparison between Fig.\ref{fig7_a_3D_M4} and Fig.\ref{fig7_b_3D_M2}, the number of EDUs is also a key factor affecting the system's SE under a certain number of receiving RRUs $L$.

\begin{figure}[!h]
  \centering
  \includegraphics[width=3.5in]{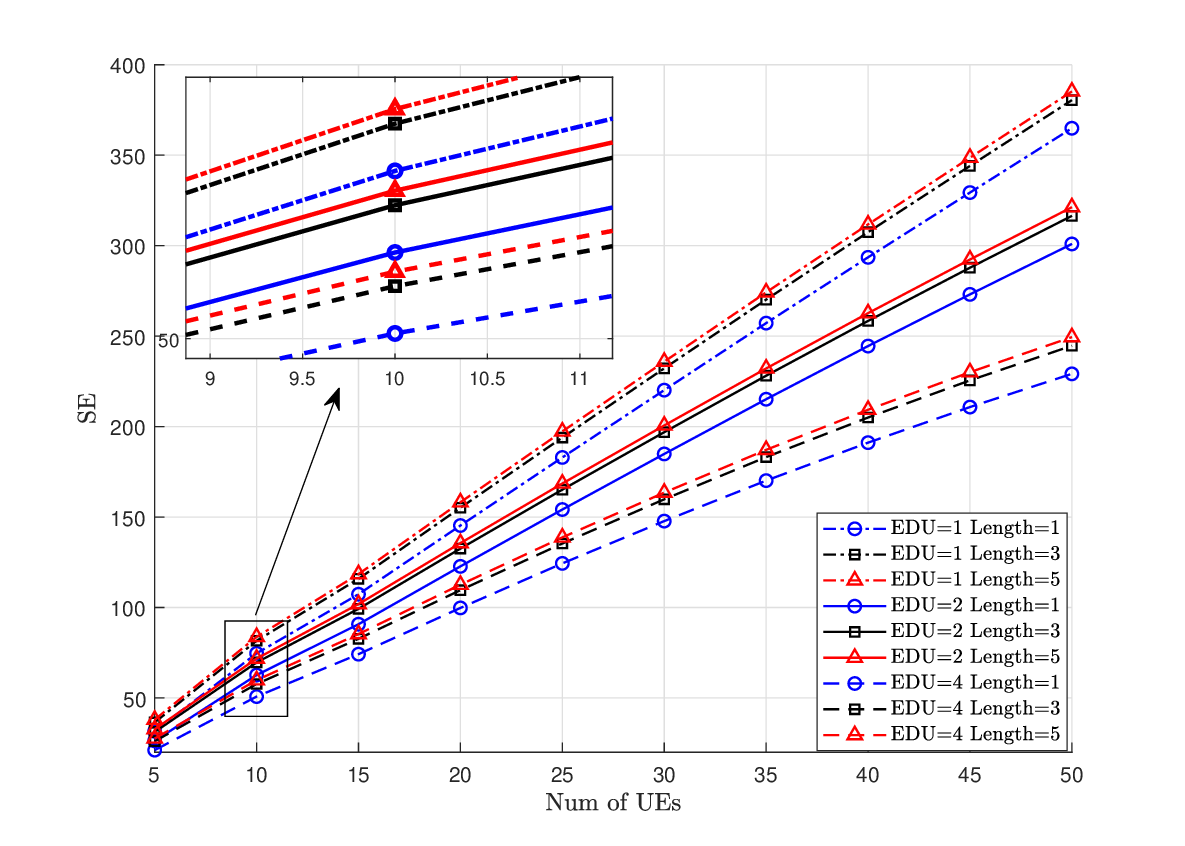}
  \caption{The system SE performance changes with the number of UEs $K$ under different block lengths $n$ and different numbers of EDUs $M$.}
  \label{fig8}
\end{figure}

The results in Fig.\ref{fig8} simulate the SE of finite block length as the number of UEs $K$ changes under different block length $n$ and various numbers of EDUs $M$. A scalable cell-free RAN system, the system exhibits macro diversity and with the combining scheme in (\ref{MMSE}), the system's SINR is relatively high. Fig.\ref{fig8} shows the block length results $n=1$, $n=3$, and $n=5$. In centralized and distributed deployment scenarios with $M=2$ and $M=4$ EDUs, the SE of finite block length increases with the block length. This also confirms that low-latency requirements are more accessible to meet in cell-free scenarios compared with point-to-point transmission.
The results in Fig.\ref{fig8} alaso shows that as the number of UEs $K$ increases, the SE of finite block length also increases approximately linearly. This is due to the large number of antenna ports and layers on the receiving side $L$ of the RRU.

\begin{figure}[!h]
  \centering
  \includegraphics[width=3.5in]{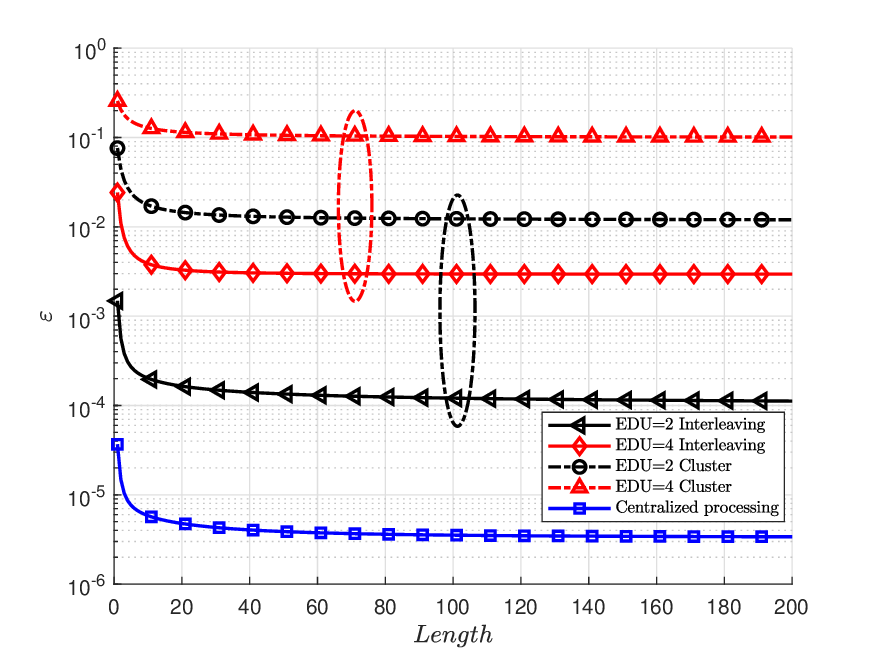}
  \caption{The system transmission error probability $\varepsilon$ changes with block length $n$ under different association strategies.}
  \label{fig9}
\end{figure}

Fig.\ref{fig9} compares the transmission error probability $\varepsilon$ with different deployment schemes as the block length $n$ changes in a system with a fixed number of RRUs  $L=100$ and $N=4$. When centralized processing is adopted, the system's reliability is the highest. As the number of EDUs $M$ increases, the transmission error probability $\varepsilon$  rises, and reliability continuously decreases. As the block length $n$ increases, the system's transmission error probability tends to a constant value. 

\begin{figure}[!h]
  \centering
  \includegraphics[width=3.5in]{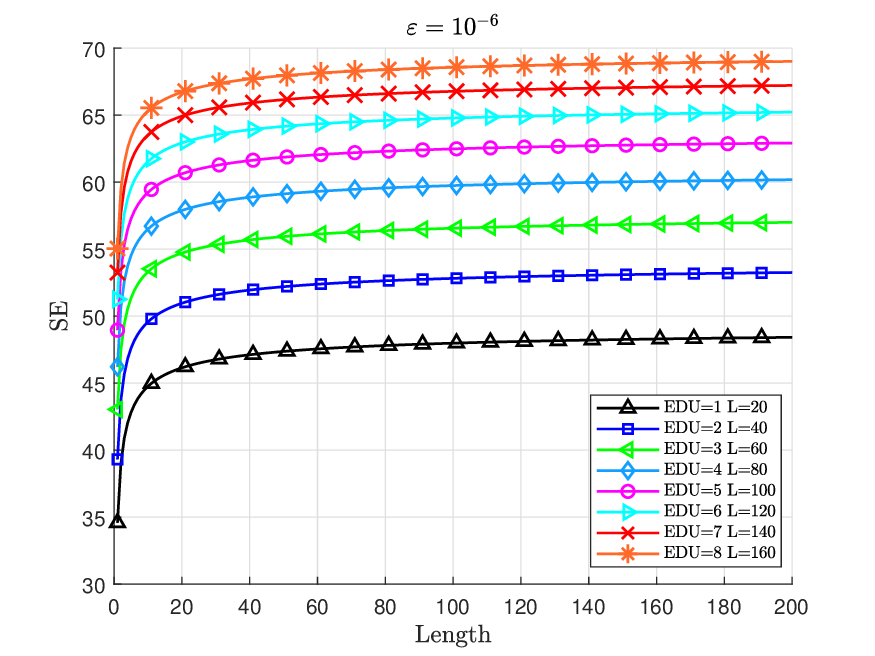}
  \caption{The system SE changes with the number of deployed EDUs $M$, where each EDU includes $20$ RRUs, all adopting the interleaving deployment, with $\varepsilon = 10^{-6}$.}
  \label{fig10}
\end{figure}

In Fig.\ref{fig10}, under the scenario of the number of UEs $K=10$ and transmission error probability of $10^{-6}$, the finite block length SE is simulated as the number of EDUs $M$ changes. It is assumed that each EDU deploys $20$ RRUs, and each RRU uses the interleaving deployment strategy. The simulation results show that with the increase in the number of EDUs $M$, the SE of finite block length continuously improves and increasingly approximates traditional channel capacity with the increase in block length. It can be seen that with the linear increase in deployed antennas, the improvement in SE of finite block lengths is not linear. 

\begin{figure}[!h]
  \centering
  \includegraphics[width=3.5in]{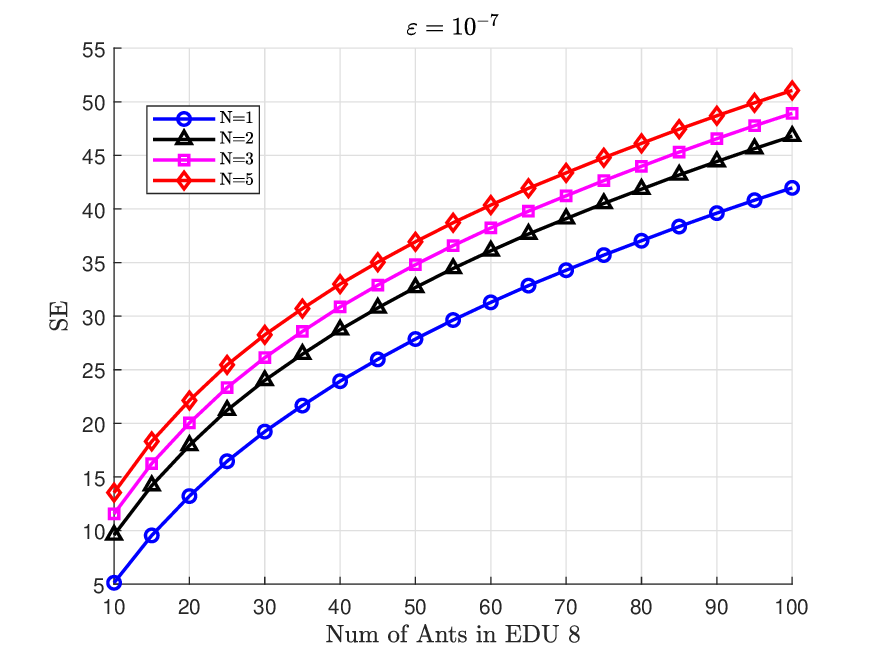}
  \caption{In the ultra-short block length scenario, the system SE changes with the number of antennas per EDU, with $\varepsilon = 10^{-7}$.}
  \label{fig11}
\end{figure}

Fig.\ref{fig11} simulates the system's SE as the number of antennas $L/M$ in the EDU changes under the transmission error probability of $10^{-7}$ in an ultra-short block length scenario. The SE of finite block lengths increases with the number of antennas in the EDU. As the block length increases, performance continues to improve.

\begin{figure}[!h]
  \centering
  \includegraphics[width=3.5in]{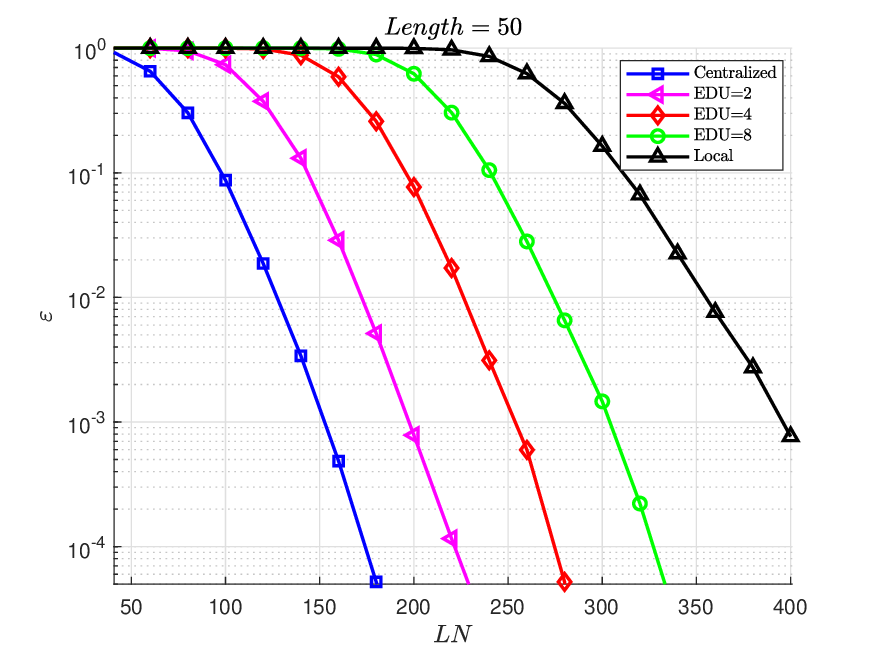}
  \caption{With $K=10$, the system transmission error probability $\varepsilon$ changes with the number of system antennas $LN$.}
  \label{fig12}
\end{figure}

Fig.\ref{fig12} simulates the system's transmission error probability $\varepsilon$ as the number of system antennas $LN$ changes in block length $n=50$ and number of UEs $K=10$. It can be seen that as the number of antennas increases, the system's transmission error probability continuously decreases. This illustrates the trade-off between scalability and reliability, and we can enhance system performance through more EDU deployments.

\begin{figure}[!h]
  \centering
  \includegraphics[width=3.5in]{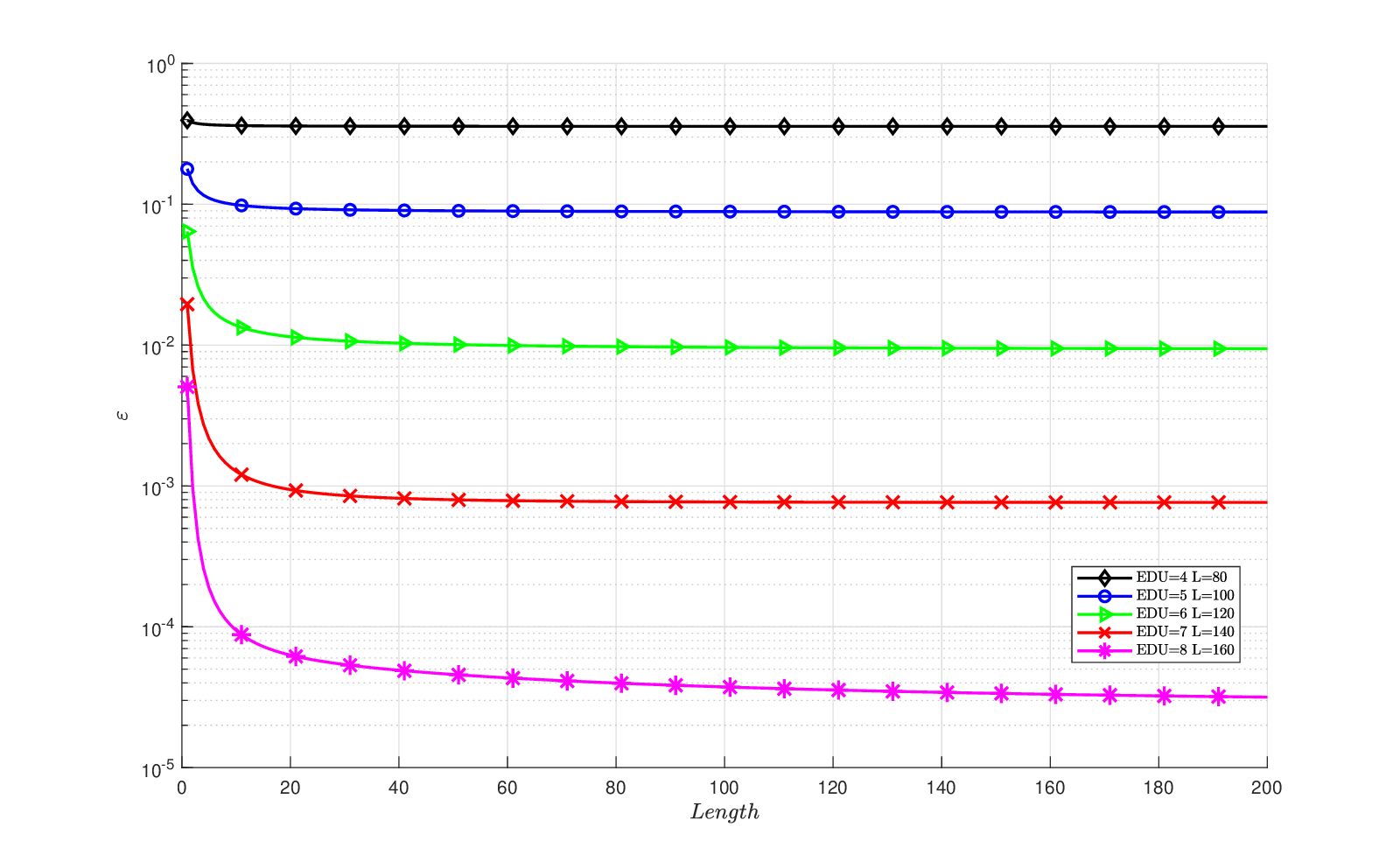}
  \caption{With $K=10$, each EDU associated with $20$ RRUs, the system transmission error probability $\varepsilon$ changes with block length $n$ under different numbers of RRUs.}
  \label{fig13}
\end{figure}

Fig.\ref{fig13} simulates the system's transmission error probability $\varepsilon$ as the block length $n$ changes with $K=10$ and each EDU associated with $20$ RRUs and deploying different numbers of EDUs. The simulation results show that as the block length $n$ increases, the system's transmission error probability $\varepsilon$ tends to a constant value.
Furthermore, with more spatial resources deployments, the system's reliability continuously improves.
Therefore, spatial resources can be exchanged for more reliability, and this performance improvement is significant.

From the above conclusions, we can see that by utilizing spatial resources, we can effectively exchange for system reliability. Assuming that the block length $n$ is directly related to the system's over-the-air transmission latency, in a scalable cell-free RAN architecture, due to the system's macro-diversity and high SINR, with the number of receiving antennas far exceeding the number of UEs, the capacity collapse caused by finite block length can be effectively alleviated.

\section{Conclusions}
In this paper, we relied on the validated space-time exchange theory in point-to-point transmission to implement uRLLC for scalable cell-free RAN systems.
The SE of a new scalable cell-free RAN with multiple EDUs was analyzed in the scenario of finite block length.
Besides, the correlation performance of RRUs under multiple EDUs was examined using a improved graph coloring algorithm for interleaving correlation, which improved system SE under block length and error performance constraints.
By deploying scalable EDUs, a balance between reliability and latency was achieved and traded off with spatial DoF, further expanding and verifying the accuracy of the distributed space-time exchange theory.

{
\bibliographystyle{IEEEtran}
\bibliography{IEEEabrv,mybibfile}
}

\ifCLASSOPTIONcaptionsoff
  \newpage
\fi

\end{document}